%% file: _INDEX.tex
\documentclass[10pt,journal,compsoc]{IEEEtran}
\ifCLASSOPTIONcompsoc
  \usepackage[nocompress]{cite}
\else
  \usepackage{cite}
\fi
\ifCLASSINFOpdf
   \usepackage[pdftex]{graphicx}
\else
   \usepackage[dvips]{graphicx}
\fi
\include{customization}

\begin{document}
	\title{Visual Exploration of Movement Relatedness for Multi-species Ecology Analysis}
	\author{Wei Li, Mathias Funk, Jasper Eikelboom, and Aarnout Brombacher}
	\IEEEtitleabstractindextext{%
		\begin{abstract}
			Advances in GPS telemetry technology have enabled analysis of animal movement in open areas. Ecologists today are utilizing modern analytic tools to study animal behaviors from large quantity of GPS coordinates. Analytic tools with automatic event extraction functionality can be used to investigate potential interactions between animals by locating relevant segments in movement trajectories. However, such automation can easily overlook the spatial, temporal, social context as well as potential data problems. To this end, this paper explores the visual presentations that also clarify the spatial-temporal contexts, social surroudings, as well as underlying data uncertainties of multi-species animal interactions. The outcome system presents the proximity-based, time-varying \textit{relatedness} between animal entities through pairwise (PW) or individual-to-group (i-G) perspectives. Focusing on the relational aspects, we employ both static depictions and animations to communicate the travelling of individuals. Our contributions are a novel visualization system that helps investigate the subtle variations of long term spatial-temporal relatedness while considering small group patterns. Our evaluation with movement ecologists shows that the system gives them quick access to valuable clues in discovering insights into multi-species movements and signs of potential interactions.
		\end{abstract} 
		
	}

	\maketitle
	
	\IEEEraisesectionheading{\section{Introduction}\label{sec:introduction}}
	Our interest in animal movement is arguably as old as the history of science itself. The earliest explorations into this by Aristotle~(384 - 322 BC) date back to 4th century BC\cite{holden_inching_2006}. With modern technologies such as sensory technology, telecommunication\cite{cagnacci_managing_2008,gor_gata_2017,hoflinger_motion_2015,qin_jiang_recognition_2004},  GIS\cite{teimouri_deriving_2018,sarkar_analyzing_2015,wang_new_2016}, and data mining\cite{li_mining_2012,li_movemine_2011}, the subject has evolved into a field where data analysis of volume entities in large open areas plays a central role in scientific discoveries\cite{nathan_movement_2008, kranstauber_movebank_2011}. As a result, scientists in the field of movement ecology are facing challenges from surging data availability and increasing demand for higher-level analytical capabilities\cite{cagnacci_animal_2010}. Today, dedicated sense-making technology has become an essential stepping stone to unlock the full wealth of ecological data at hand\cite{spretke_exploration_2011}. 
	
	Spatial movement is an important feature in finding ecological patterns\cite{holden_inching_2006, nathan_emerging_2008, westley_peter_a._h._collective_2018}. As a recent trend in movement ecology, ecological insights into smaller stages has received proliferated attention\cite{holyoak_trends_2008}, particularly the individual level analysis of movements\cite{calabrese_disentangling_2018,giuggioli_stigmergy_2013,polansky_framework_2011,strandburg-peshkin_inferring_2018,torney_inferring_2018}. In this category, cross-species interactions (antagonistic interactions like host - parasite, predator – prey and plant – herbivore\cite{pires_interaction_2012, hagen_biodiversity_2012}, for instance) are important because they are typically studied at a zoomed group scale, where individual behaviors are investigated. 
	
	

	
	Moreover, movement ecologists try to look beyond interaction instances alone. They wish to interpret the higher level meaning or motivation behind certain behaviors as a long-term pursuit\cite{slingsby_exploratory_2016, westley_peter_a._h._collective_2018}. As phone data can reliably predict human social ties by their propinquity\cite{eagle_inferring_2009}, the insights into the relationship between animals might be useful to collect evidence for investigating the less approachable motivations and meanings behind certain movements. This can be facilitated by visualizing movements that sufficiently express the details that relate to not only individuals in isolation but also the potential influence between adjacent group of multi-species entities. 
	
	
	To this end, we design an interactive visualization system that addresses the above concerns and help ecologist in the tasks of finding interesting patterns of animals' relations. The implementation of the interactive system supports movement analyses through the perspective of relatedness, which concerns social intimacy or potential predatory threats in wild environments. It is determined by the general spatial proximity that varies unstably in longer periods. 
	
	
	We validate the system's usefulness through use cases and expert evaluations. Positive results are received. This paper posits its contribution as an exploration characterized by:
	\begin{itemize}
		\item a simplified, animation-based visual vocabulary designed to facilitate the communication of movements and time varying relatedness,
		\item a visualization method to support pairwise and individual-to-group comparison of animal's spatial situations, and
		\item including uncertainty awareness in the study of animal movement relations.
	\end{itemize}


	%
	%
	
	\section
	{Related Work}
	\label{sec:related-work}
	
	\subsection
	{Visualization in Movement Ecology}
	\label{subsec:visualization-movement-ecology}
	Movements are usually referred as locomotive movements in the field of movement ecology. Location changes are useful clues to reveal ecological patterns in the problems such as resource use\cite{roshier_animal_2008}, population dynamics\cite{cagnacci_animal_2010}, and climate influence\cite{ferreira_birdvis_2011} between individuals, groups, or species. Many ecologists have already known the value of visualization in their ecological research. Generic visualization tools (e.g. Movebank\cite{kranstauber_movebank_2011},  AMV\cite{kavathekar_introducing_2013}, Env-DATA\cite{dodge_environmental-data_2013}) are employed to support common tasks like trajectory plotting and multivariate filtering. As they cover a wide range of species and data types for many research projects, these tools are enough for basic movement analyses. 
	
	
	However, some research tasks require analytic aggregation capabilities. Drosophigator\cite{seebacher_visual_2018} uses statistics from heterogeneous data sources to generate visualized predictions of the spread of invasive species. Xavier\etal\cite{xavier_exploratory_2014} integrates environment data to study the connectivity (a technical term in ecology indicating the degree of environmental variables affecting its inhabitants in an area\cite{taylor_connectivity_1993,baguette_landscape_2007,lima_towards_1996}) of landscape characteristics and animal behaviors. Konzack\etal\cite{konzack_visual_2018} analyze the migratory trajectories to recognize the stopovers among gulls' movements.
	
	Visual design is also important to communicate the aggregated results that relates to the exact domain problem. Slingsby\etal\cite{slingsby_exploratory_2016} discusses the design choices of visual encoding in ecology visualization, suggesting that the use of visual language needs to convey the "ecological meaning" to support the research context. Spretke\etal\cite{spretke_exploration_2011} conceives the "enrichment" method as a way to enhance analytical reasoning with the visual elements of attributes like speed, distance, duration in the geographical context, allowing quick hypothesis iterations on local subsets. 
	
	Aggregation of attributes of individuals might be useful to understand the movement behavior, but behaviors is better explained in a context where influence of peers and surroundings are considered\cite{nathan_movement_2008}. Since the interest in entity level behaviors is rising\cite{holyoak_trends_2008}, investigating behaviors concerning the mutual influence of multiple entities can be a worthy starting point. 
	
	%
	%
	%
	\subsection{Trajectory Analysis}
	\label{subsec:trajectory-analysis}
	Manually searching for patterns in the long, twisted, and sometimes cluttered movement trajectories can be daunting. This makes event detection algorithms necessary as they alleviate the cognitive load for experts. The execution of automatic event detection usually requires defining a set of parameters, such as time window, speed, heading, and mutual distance, to narrow down the search space to a subset of trajectory fragments\cite{bak_scalable_2012, andrienko_event-based_2011,siqueira_discovering_2011,andrienko_uncovering_2008}. The detection outcomes are usually visualized on top of the movement trajectories with reference to the original geographic context. As the detection processes are sensitive to the subject animal and landscape context \cite{siqueira_discovering_2011}, we need flexibility in a visualization tool to cope with the distinct characteristics of movements for a convincing result\cite{bak_scalable_2012}. For example, Andrienko\etal\cite{andrienko_uncovering_2008} suggested a bottom-up approach where the detected elementary interactions (a concept derived from Bertin's elementary level of analysis\cite{bertin_semiology_2010}, meaning "particular instances of interaction between individual objects") are used as key clues to understand group level patterns. Bak\etal\cite{bak_scalable_2012} proposed method to boost the performance in event detection at larger scale. The extra performance gain can thus be allocated to support interactive parameter input, through which the visual feedback of outcomes makes an essential part of the interactive loop to guide the next iteration. Bak\etal also mentions the classification of four types of higher level encounter patterns, which seems to be a continuation from Andrienko\etal\cite{andrienko_uncovering_2008}'s advocacy of characterizing elementary patterns.
	
	Movement interactions are multifaceted. A visual feedback for adaption and fine-tuning of the analytic system is indispensable to extract interactions. Also, automatic techniques mostly solve lower level tasks such as matching route similarity or spatial-temporal closeness between trajectories. Recognizing general patterns and questioning with alternative assumptions are still a job of human expertise. Thus, we need to keep a open mind to the machine results but, in the meanwhile, expose more visual details for domain judgment. 
	
	\subsection
	{Visualization of Spatial Temporal Movements}
	Movement traces are usually placed on a 2D map as discrete points\cite{slingsby_exploratory_2016} or linked trajectories\cite{konzack_visual_2018,shamoun-baranes_analysis_2012}. However, consideration of both spatial and temporal changes are necessary to avoid false identification of collocation (position overlap at different time period). There are several approaches to achieve this.
	
	A common treatment here is the Space Time Cube (STC)\cite{hedley_hagerstrand_1999,gatalsky_interactive_2004,kraak_space_2003}, which projects the temporal dimension in the z-aixs of a 3D view. In STC, real collocation of two entities are depicted as the neighboring points in a 3D space. But it makes visualization work prone to problems like loss of perspective and obfuscation\cite{walsh_temporal-geospatial_2016,amini_impact_2015}. Subjects that travels in the z-axis can also be another issue\cite{andrienko_clustering_2018}. A workaround without using 3D space can be found in AMV\cite{kavathekar_introducing_2013}. It confines trajectories to the local duration and presents relative movements by removing distractions of trajectory parts that are temporally distant. But the fine details of proximity variations in the selected duration is not supported. Alternatively, abstracting  movements from their geo-spatial context, as explored by Crnovrsanin\etal\cite{crnovrsanin_proximity-based_2009}, is also a possible way to clarify the subjects' spatial temporal relations.
	
	In sum, we found visualization facilitation in the wildlife behavior analysis is non-trivial, especially when support of time space relatedness remains an open gap. This motivates new solutions with flexible interactivity for quick selection, navigation, and visual adjustments \cite{andrienko_visual_2013-6} to assist the domain research.
	
	\section
	{Context and Requirements}
	\label{sec:context-and-requirements}
	In this section, we describe the basic setup of the domain research including the expert collaborators, metadata, apparatus (for data collection), and domain requirements. 
	
	\subsection
	{Project Background}
	\label{subsec:project-background}
	The domain problem targets at a group of multi-species, free-roaming animals in a South African nature reserve. Researchers' primary interest lies in the behavior along space/time variations. Instead of analyzing relationships with natural landscape, experts needs visual insights into behaviors of individual animals and how they would influence each other, which is also a valuable compensation to their current tool sets. Individual interactions are also potential to later applications such as to analyze other species, or even human social interactions.
	
	Two ecologists are invited as domain experts (E1 and E2). They both have long interests and experience in animal movements. E1 has a background in movement ecology, spatial analysis. E2 also works in quantitative ecology and environmental sciences. They both conduct quantitative data analysis with R and use field-specific packages to plot the movements either by spatial attributes~(map) or numeric attributes (bar chart or line chart). However, they both find their current tools limiting. For example, both of them complained about the lack of support in converting the multidimensionality of movement trajectories into useful information. E2 also mentioned that the analysis becomes challenging when facing multiple individuals regarding their temporal-spatial variations.
	
	\subsection
	{Data Description}
\begin{figure}[h]
		\centering
		\includegraphics[width=0.5\linewidth]{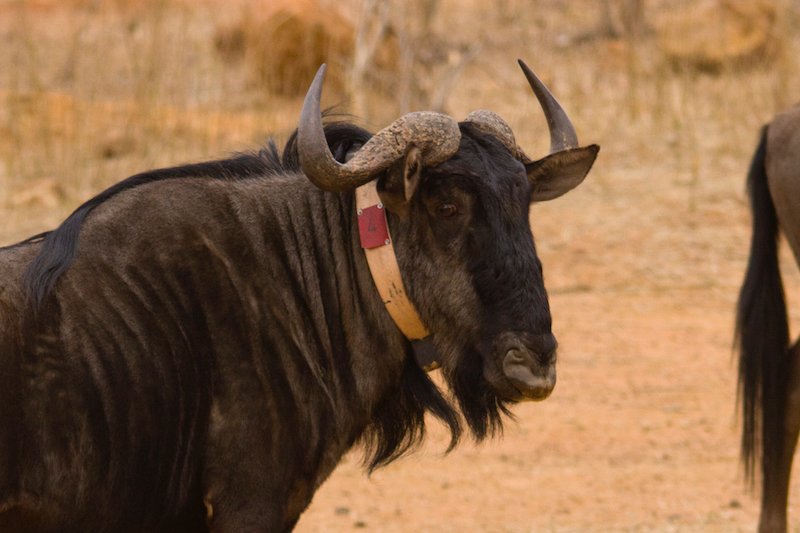}
		\caption{GPS Collar: The device is programmed to record the GPS coordinates repeatedly with a fixed time interval.}
		\label{fig:collar-tag}
	\end{figure}

	For data collection, ecologists attach GPS collars\cite{handcock_monitoring_2009} to each of the 25 sampled animals, cf. \autoref{fig:collar-tag}. The subjects are consisted of 5 lions, 10 wildebeests and 10 zebras. The entire data collection lasted for roughly 30 months, which captures periodic changes of climate though different seasons. To ensure sufficient battery life, collar tag sensors were programmed to obtain and store GPS coordinates by every two hours. 
	
	Due to various conditions in climate and land features, the quality of tracking data suffered from  unpredictable wild circumstances\cite{hurford_gps_2009}. Cause of quality compromise can be any incident such as physical impact, wet situations on rainy days or water areas, or poor signal reception. To some extent, the collected data are often compromised with one or more of the following issues: a) unrealistic values, i.e. two subsequent data points have impossibly large position difference in between which is evidently an error, and b) missing points, i.e. failure to record a data point at the programmed time. Therefore, pre-processing before analysis is needed\cite{bjorneraas_screening_2010}. With their expertise, domain experts proceeded with the removal operations to deal with the first type of error (a), resulting in unstable trajectories with irregular gaps of varying length on it. They also trimmed off the unusable parts at the beginning or ending of every animal tracking, making each trajectory starts and ends at different time of its own.
	
	\subsection
	{Requirements}
	\label{subsec:requirements}
	Integrating a visual approach into existing analytic workflow requires shared understanding from both ecology and visual analysis. Through six months' exchanges with the domain experts, we have devised five design requirements as guidelines to add new capabilities to the current analysis. We begin with draft design proposals in the form of low-fi visual sketches, which inspire nuanced discussions to pin-point their most prominent concerns. To communicate more complex effects, we built iterations of animations with Processing\cite{reas_processing_2007} and evaluate of their usefulness in each version. There are also co-design sessions during which they take a more active role to freely propose ideas though paper sketches and explain the intended functionalities. To test for likely insights and also unexpected misinterpretations, qualitative surveys were also conducted among an expanded user group (7 people) of both experts and non-experts. We used structured questionnaires to prioritize the requirements and balance conflicting directions. With constant refinements, we conclude that the visualization design should facilitate movement ecology research by addressing the requirements in the following aspects:
	
	
	\emph{R1 - The ability to intuitively visualize spatial temporal movements in a simple and straightforward manner.} 
	
	The experts understand the challenges introduced by the multidimensionalilty of collected data. But their experience with interactive visualization tools with sophisticated functionalities is limited. Steeper learning curves can be worrying as they have developed existing habits in thinking with movements. So a design with simple and intuitive depictions would be preferred. They understand the value and efficiency of visual exploration in finding implicit patterns which can be easily missed otherwise. With minimum extra explanation, it should support them in confirming newly discovered visual knowledge with fine data details and facilitating model building with existing analytical skills. 
	
	\emph{R2 - Interactivity and navigation into spatial temporal space efficiently.}
	
	Considering a 2.5 years long time frame of the project, browsing the global timeline means navigating roughly 4000 data points for every 2 weeks of tracking time. Additionally, the patterns can exist in any level of time scopes, e.g. movement difference in days and nights, weeks and even seasons. Granular time scopes should be implemented as movement patterns are sensitive to the observation time frame. Interactively defining time point and temporal selection with flexible scope length will give them the autonomy to explore for a suitable length for particular research task.
	
	\emph{R3 - Ability to visualize the movement with awareness and tolerance of inconsistent data quality.}
	
	As there are known issues in the data quality, the visualization pipeline should include the ability to produce smooth, consistent visual results. However, instead of ignoring the problems in data, the involved uncertainties should be truthfully presented so that user could easily distinguish whether the visualized outcome is based on solid trustworthy data or questionable delusion. A visual distinction should accommodate the difference clearly and allow for context-dependent judgments.
	
	\emph{R4 - Comparing movements between different species.}
	
	Same species share behavioral similarities. And the opposite is true for different species. But extent of difference varies between sets of comparisons. For instance, both being herbivores, the wildebeests and zebras can find more commonalities, while lions, as predators, may behave differently than herbivores. To explore species-relevant aspects in movement behaviors, analysis with species awareness is needed. The visualization system should allow them to visually disintegrate the behavior differences, making the anomalies instantly distinguishable. 
	
	\emph{R5 - A focus on the relationship between animals.}
	
	Movement behavior is a complex problem which can be influenced by many environmental factors. It is also not surprising that animals would influence each other's movements in more or less subtle ways from slight movement deviations to social interactions. Understanding how animals would influence or interact with their movements by both intra-species or inter-species means can contribute to valuable ecological insights. In our case, the queries into the dynamics of how animals are closely bounded to each other and whether such bounding is consistent should be facilitated.
	
	As a side note, the requirement list follows a hierarchical structure where meeting the later, higher level ones involve solving basic ones prior to them.
	
	\begin{figure*}[h!]
		\centering
		\includegraphics[width=.8\linewidth]{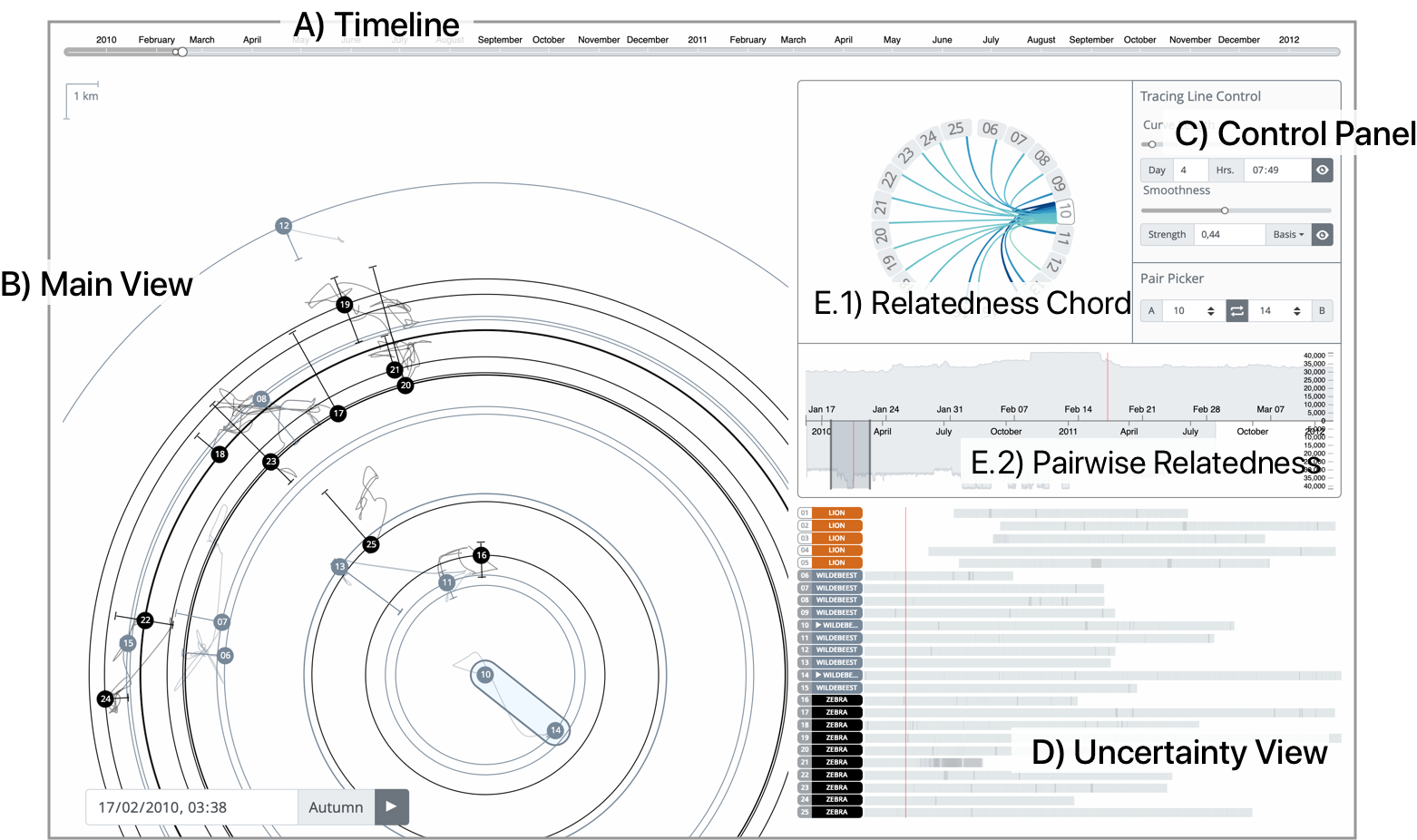}
		\caption{Interface Overview: A) Timeline: global timeline to indicate and cast universal time change. B) Main View: a 2D space to display geo-spatial relations. C) Control Panel: control widgets to fine tune parameters for animation or pick animal pairs. D) Uncertainty View: an overview of collective data stability and availability. E.1) Relatedness Chord: A chord diagram to display relatedness between multiple entities. E.2) Pairwise Relatedness: Interactive line chart for long term pairwise relatedness display.}
		\label{fig:overview}
	\end{figure*}
	
	\section
	{Design Rationale} 
	\label{sec:design-rationale}
	In search for proper methods and to convert the requirements to executable guidelines, we present a few conceptual elaborations as cornerstones to support the design. 
	
	\subsection
	{Parametric Moving Trace Line}
	\label{subsec:parametric-moving-trace-line}
	
	As aforementioned, context adaptability is important in movement analysis. Parameterized visualization is an effort to support this by controlling local variables, such as species, speed, or group size, to address the problem of concern. In our case, larger time interval size is used to record movement locations. Straight lines connecting location points would appear cluttered with angular shapes and presents little clue to anticipate the movements between locations. Regarding this issue, trajectory smoothing\cite{sacha_dynamic_2017} can be helpful as the visualized path can help to build visual heuristics to makes potential behavior patterns easier to uncover and trajectories of different individual more visually distinguishable. The smoothing also produces other effects that will benefit the analysis. For example, the smoothing will make sharper turns with slower animals (ones with shorter distance between with consecutive points), while the opposite applies to faster moving animals. Domain experts can also use a middle point in the curve to adjust their estimations of the heading if necessary.
	
	A decision upon the right amount of smoothness and trajectory shape requires careful deliberation and calibration. To support easy reconfiguration, parameter setting with instant visual feedback can improve the productivity of iterative decisions. This helps domain experts to understand the effect of each parameter to facilitate easier interpretation of trajectory lines regarding~\textbf{R1} and creates adaptability for different analytic scenarios. We understand that the domain experts may occasionally wish to fact-check the bare data points, therefore smoothing should be implemented with an option to be temporally turned off to make explicit locations points visible.
	
	\subsection
	{Defining Spatial Temporal Relatedness}
	\label{subsec:defining-spatial-temporal-relatedness}
	The inherent entanglement of space and time (defined as \textit{spatio-temporal dependency} in GIS) is a pivotal challenge in movement analysis\cite{demsar_analysis_2015}. Observations in the spatial domain without consideration of temporal stability may result in unreliable or even false positive discoveries\cite{andrienko_visual_2013-6}. Assisting domain experts to think temporally is essential\cite{andrienko_space_2010}.
	
	To find potential interactions between animals, the spatial proximity should be discussed within the relevant duration context. Based on this principle, relatedness is a concept that describes relationships with spatial-temporal reference. To distinguish relatedness from proximity, relatedness concerns the distance variation in a time range, whereas the proximity only concerns distance in static a time point. Thus, the relatedness can be treated as summary of physical proximity through time. 
	
	We employ two basic modes for inspecting how related animals are bounded to each other --  the \textit{pairwise} ~(\textbf{PW}) relatedness and the \textit{individual-to-group}~(\textbf{i-G}) relatedness. The pairwise approach takes two entities as input and displays their dynamic relatedness variations. The i-G approach takes a focal entity and displays how closely it is connected to the rest of the animal group over time. These two modes compensate each other for different tasks. 
	
	Pairwise relatedness~($\mathit{P_{ij}(t),\; t \in T, \; i,j \in A}$) is a time dependent scalar value describes physical proximity ($\mathit{d_{ij}}$) comparing to the maximum distance ($\mathit{M}$) bounded in the captured area, i.e. $\mathit{P_{ij} = M - d_{ij}}$. Here, $\mathit{T}$ represents all the possible states in the global timeline, while $\mathit{i}$ and $\mathit{j}$ are two elements from all the animals ($\mathit{A}$). Relatedness of one versus multiple targets enables comparison between more candidates and is more complex than pairwise mode. Because the collective pattern of i-G relatedness $\mathit{\,G_{x}(t)}$ is  contextually understood as a qualitative, multivariate pattern which considers explicit movement trajectory, relative proximity, and relatedness trend which involves constant changing features among multiple targets. To ease communication, we use a different notation to represent: $\mathit{G_{x}(t) = \{ P_{xa} (t) \;| \; a, x \in A \; and \; x\neq a, \; t = \{t_{1}, t_{2}, ..., t_{R}\} \} }$, where $\mathit{x}$ is the focal animal (the one to be compared with the rest of the group) and a time range from $\mathit{\;t_{1}\;}$ to $\mathit{\;t_{R}\;}$ is considered. To deal with the intricacies that are difficult to be summarized as a quantitative measurement, we implemented a combination of visual components to delineate the implicit patterns and variations.
	
	
	\subsection
	{Uncertainty Awareness}
	\label{subsec:uncertainty-awareness}
	The uncertainty awareness is believed to be useful in improving decision-making\cite{riveiro_effects_2014, wunderlich_visualization_2017}. Communicating uncertainty is important in our case because it can false signal the absence of one entity in a mutual relationship, which could mislead the experts' judgment and indicate a termination of related session. To avoid this, we take a series of measures: Firstly, we perform linear interpolation to fill the missing gaps to ensures steady data flow for the well-functioning of relatedness derivation model. Secondly, we treat the interpolated data points differently by labeling and measuring their reliability. This is realized by assigning a special Boolean label, i.e. interpolated or null, and a measurement of uncertainty extent. The measurement is computed as follows: given the current index $\mathit{i}$ and index range of consecutive missing data points $\mathit{[c, c']}$ ($\mathit{i \in [c, c], ~~i,c,c' \in N} $), the degree of uncertainty $\mathit{U}$  can be formulated as: $\mathit{U = min(|i - c|, |i - c'|)}$, i.e. the uncertainty in current data point $\mathit{i}$ will be determined by the index distance to the nearest (temporally) reliable data point.
	
	Thus, the uncertainty information prepared to be visualized with different levels of awareness(\textbf{R3}). The experts can make judgment on the credibility and integrity of a pattern by also referencing the visualized uncertainty\cite{sacha_role_2016}. 
	
	\section
	{System Description}
	\label{sec:system-description}
	We implemented our visualization system in an interactive, web-based application with D3.js. It consists of a main view,  peripheral views, and control areas, see \autoref{fig:overview}. The functionalities are results from \secref{subsec:requirements} and \secref{sec:design-rationale}. \textbf{Main area} (\autoref{fig:overview}~B) displays movements and trace lines in relation to their original geographical patterns. \textbf{Timeline} (\autoref{fig:overview}~A) is a time reader and controller to set the global "current" time and indicate the length of covered duration, which is shown as the red line along the center (\autoref{fig:overview}~C). Trace line adjustments and animal pair selector is placed in the control panel. Uncertainty in the data are indicated separately by each animal along the time progression (\autoref{fig:overview}~D). Relatedness measurements can be found in \autoref{fig:overview}~E1, E2. We expand with more details in the rest of this section.
	
	\subsection
	{Movement Encoding}
	Moving animals are represented as animating locations, each with an ID and a species color. Species are colored to resemble the animal' natural appearance: orange for lion, cool gray for wildebeest, and black for zebra.
	
	Each animal entity draws a trace line with the same color of itself, the length of which corresponds to the global duration. As time coverage is the same for every trace line, drastic movements (bigger distance between steps) will appear longer than sedentary animals, creating a contrast that enables comparison between animal individuals by its movements intensity. Thus, outlier movements can stand out more clearly.
	
	\begin{figure}[h]
		\centering
		\includegraphics[width=\linewidth]{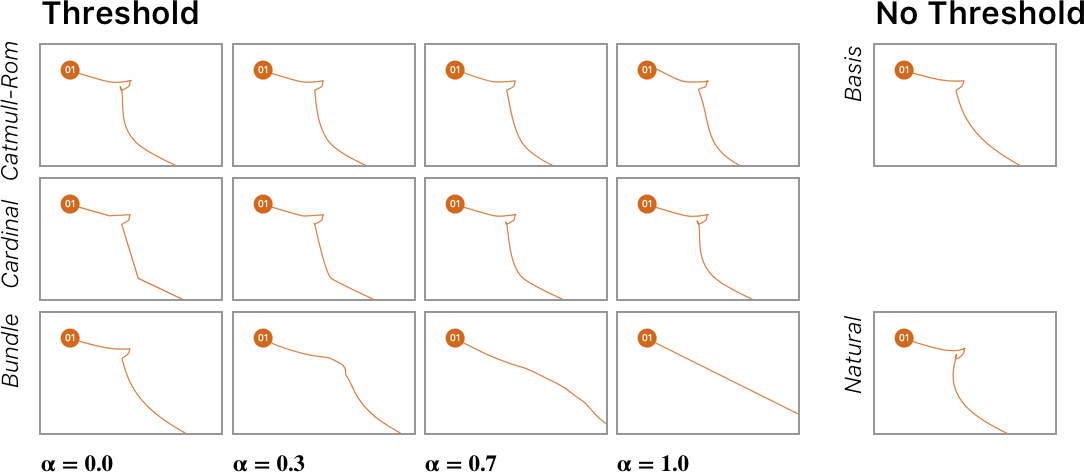}
		\caption{Curve Adjustments \& Settings: Experts can leverage their domain knowledge to tweak smoothing functions and smoothness threshold for the desired result.}
		\label{fig:curve-parameters}
	\end{figure}
	
	To improve the readability of movement trajectories (as explained in \secref{subsec:parametric-moving-trace-line}), we applied parametric smoothing to the trace lines. The technique is partially inspired by Sacha\etal's trajectory simplification~\cite{sacha_dynamic_2017}, we integrated commonly used methods like cubic basis spline, natural cubic spline,  straightened cubic basis spline based on Holten's edge bundles~\cite{holten_hierarchical_2006}, Cardinal spline, Catmull-Rom spline with D3.js' curve interpolation functions. This flexibility in configuration, cf. \autoref{fig:curve-parameters} can be fine-tuned by mode or intensity. Here, the mode determines the type of underpinning smoothing function and the intensity, specified by a linear $\alpha$ value between 0 and 1, offers the option to adjust the amount of smoothness: $\alpha = 0$ means no smoothing at all and $\alpha = 1$ means smoothest. The goal is to give domain experts extensive free options to pinpoint the right parameter to draw trajectories that caters to their analysis scenarios.
	
	\subsection
	{Uncertainty Encoding}
	\begin{figure}[h]
		\includegraphics[width=.9\linewidth]{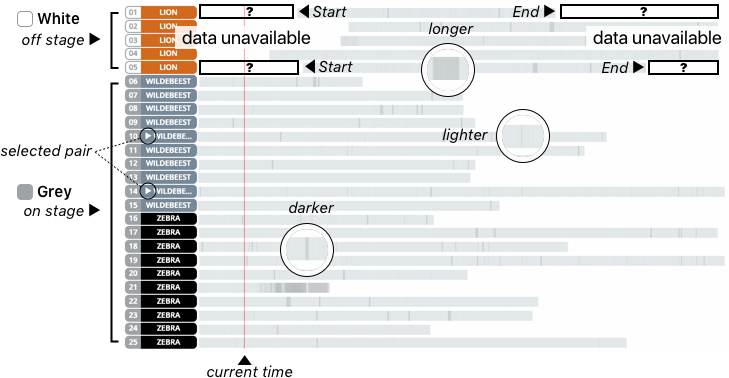}
		\caption{Uncertainty View: displaying data quality issues with temporal context. The view provides a sense of whether the data quality can be trusted around current point in time. Labels on the right show whether the animal's data is currently available to be displayed on main view to the left. The "$\triangleright$" signs indicate the selected animal pair for pairwise relatedness analysis. Visual clutter can be mitigated by clicking on the label to hide certain animal subsets.}
		\label{fig:temporal-uncertainty-view}
	\end{figure}
	
	Uncertainties can be visualized following the temporal axis, cf.~\autoref{fig:temporal-uncertainty-view}. Small horizontal heatmaps at the bottom right of the screen are drawn to inform data issues in three aspects: 1) the start and end time of available data,  2) the general distribution, and 3) the degree of uncertainty (by depth of color), cf.~\secref{subsec:uncertainty-awareness}. The time context is useful to guide expert to skip certain segments by informing where to expect reliable data.
	
	\begin{figure}[h]
		\centering
		\includegraphics[width=.9\linewidth]{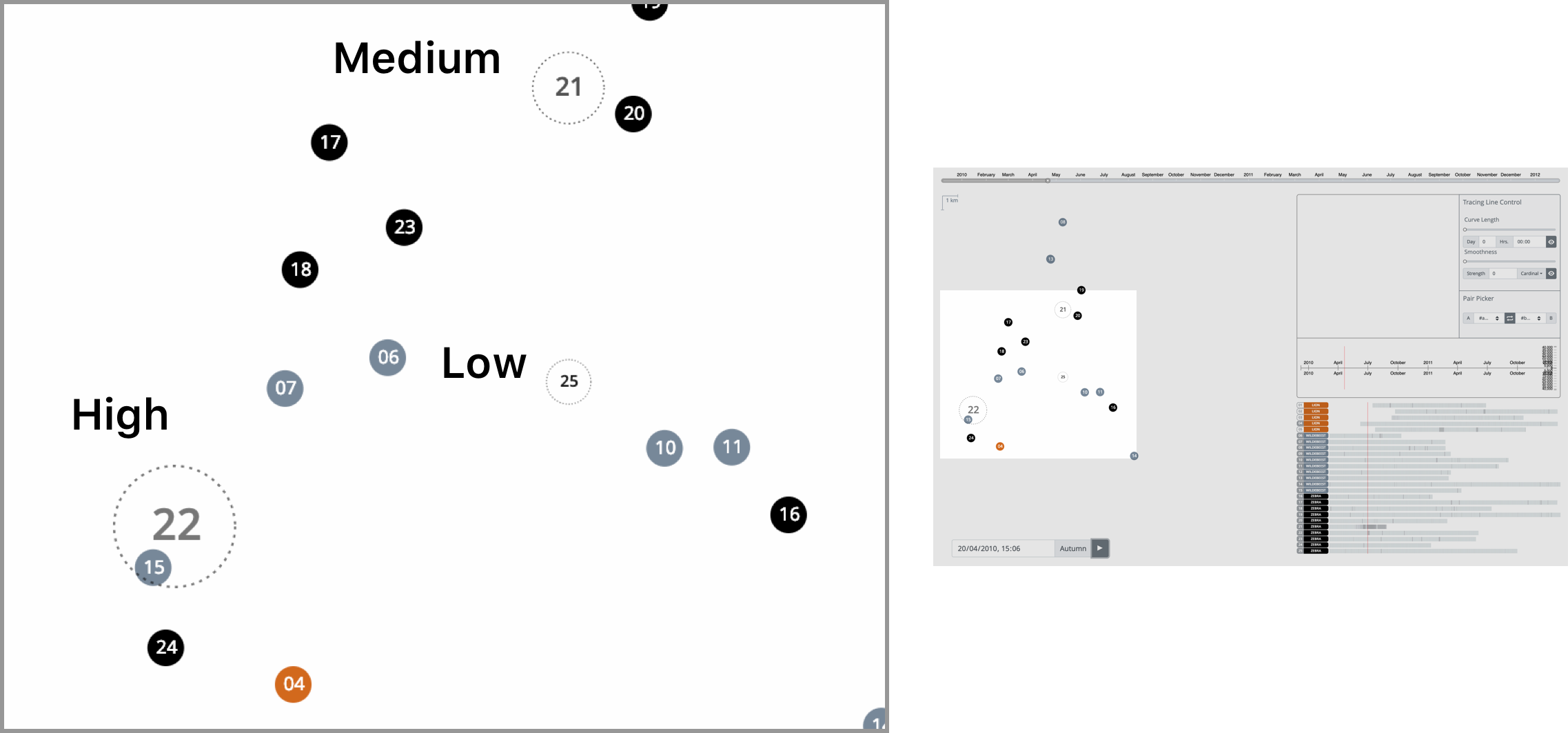}
		\caption{Uncertainty in Spatial Movements: Empty circles are shown to tell the end of  available data. A circle with expanding area indicates movements are visualized with diminishing accuracy.}
		\label{fig:locational-uncertainty}
	\end{figure}
	
	In the spatial domain, moving entities will change both appearance and size once uncertainty happened in the data, cf.~\autoref{fig:locational-uncertainty}. Filled circles change to dashed outlines, expanding their sizes to indicate a dilution of positional accuracy. Its opacity also decreases along with the size increase, telling the viewer that the system is unsure about exact location of current animal. 
	
	Expert can leverage both display methods to avoid risky interpretations and apply self-disretions with their domain knowledge. 
	
	\subsection
	{Relatedness Encoding}
	\label{subsec:relatedness-encoding}
	The relatedness can be understood through different setups. We describe two modes to treat them respectively: the relatedness between two individuals (pairwise) and one individual comparing to a group of the rest (i-G). 
	
	\begin{figure}[h]
		\centering
		\includegraphics[width=.9\linewidth]{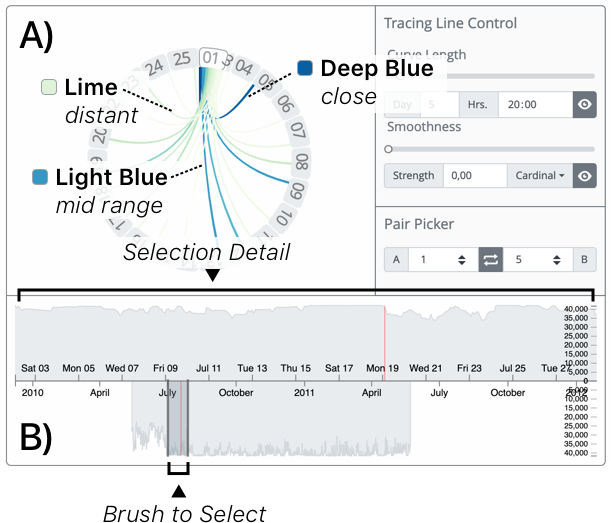}
		\caption{Pairwise relatedness in two views: A) Chord diagram with each ribbon for any possible pair in the given time range; B) Interactive line chart with separated brush-to-zoom functionality.}
		\label{fig:pairwise-relatedness}
	\end{figure}
	
	Granular variation of $\mathit{P_{ij}(t)}$ is plotted with a line chart. Overview-and-zoom functionality\cite{cockburn_review_2008} is needed here considering the amount of details in 2.5 years of data points. So we implemented a vertically mirrored line chart below it with different purposes for each half – the bottom one can be brushed to select the time range and the top side displays zoomed details of the brushed area., cf. B) in \autoref{fig:pairwise-relatedness}.
	
	The chord diagram can provide an overview of relatedness of all possible pairs being presented in the main view. Ones with higher relatedness are drawn in bolder and more saturated ribbons while lighter appearance applied for the lower relatedness, cf. A)  in \autoref{fig:pairwise-relatedness}. The delineation reacts differently to a duration and fixed time point. If a duration is selected, the system calculates averaged proximities of all intervals in the range to determine the ribbons' appearance: $\mathit{\overline{P}_{ij} = \frac{1}{R} \cdot \sum_{s = 1}^{R} P_{ij}(t_{s})}$. This approach is very similar to the pairwise mode but more aggregated. Experts can brush and drag to tweak the duration length. Such operations are useful to answer questions like "Were the animals' movements more clustered (related) during the past eight hours? Were they the same for last two days?". It is a simpler and more straightforward way to search for patterns without looking at the spatial changes in the main view.
	
\begin{figure}[h]
		\centering
		\includegraphics[width=.9\linewidth]{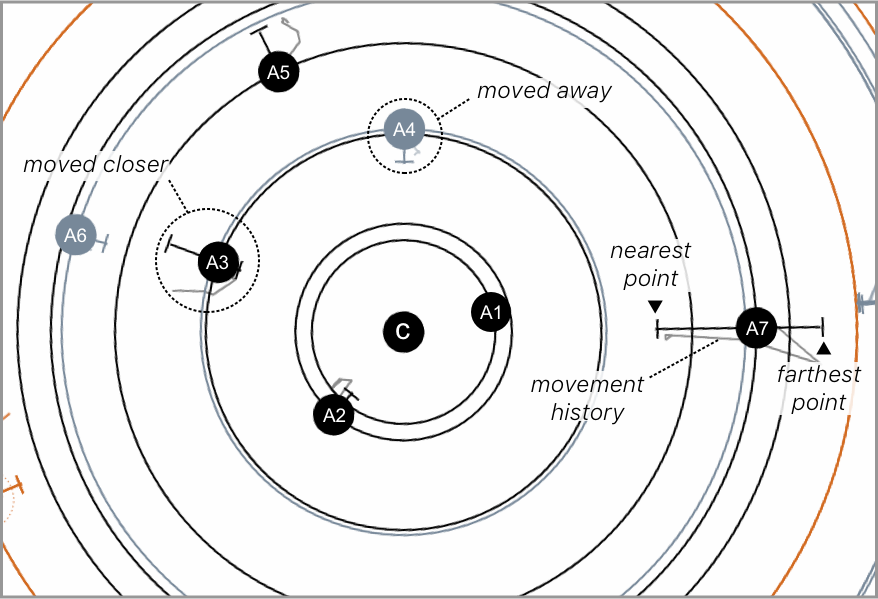}
		\caption{i-G Relatedness in Main View: Animal movements are displayed in relation to the focal animal C. Their relatedness in the duration is aggregated and represented by capped lines which ends are used to indicate the nearest and farthest distance to the focal animal. Here we can see that A3 has moved closer to C and A4 has become more distant judging by the overall trend. Despite located at different directions to C, the distance comparison between A4 and A3 is still intuitive and apparent.}
		\label{fig:ig-relatedness}
	\end{figure}

	Geographical pattern of the i-G relatedness ($\mathit{\,G_{x}(t)}\,$) is displayed in the main view where spatial distribution and social context are sensitive aspects. The individuals can be focused by clicking on its circle in the main view. The interaction halts any ongoing animations and creates an array of concentric circles. Relations of animals concerning the focal animal can be visually examined~\autoref{fig:ig-relatedness}. Radii of the circles indicate the spatial proximity to the animal: $\mathit{r = P_{xa}}$. The circles sort proximity of animals with scattered distances and moving trends (came closer or went further) into an egocentric diagram where unidirectional comparison of proximity is possible. Based on this, proximities of current time is easy to tell by circle sizes. The less explicit moving trends, however, are illustrated by the trace lines and relatedness error bars. Here, the former corresponds to the movement trajectory of the duration and the later is a depiction to analyze the trend of relatedness: line caps on both ends of the error bars are determined by the maximum and minimum values of relative proximity in the duration. Thus, whether an entity is drawing closer or moving farther can be read from the negative/positive sum of relatedness, which is derived by comparing length of error bars from the inside (positive relatedness) and outside (negative relatedness) of the proximity circle. For instance, ones with much larger outer length suggests the underlying animal spend most of the recent time in places more distant to the focal animal than its current location, which means it tends to move farther considering the past period. The trace line here is to verify the judgment on the trend of relatedness change with exact movement details. 
	
	\section
	{Use Case}
	\label{sec:use-case}
	We tested the visualization system with domain experts to validate its usefulness in practical environments. Specifically, the system is hosted online as a web application, and we recorded two sessions of screen interactions and verbal communications remotely via Skype. No specific tasks were given during the experiment. Experts are encouraged to explain their reasoning following the think-aloud protocol~\cite{boren_thinking_2000}, and support the explanation with domain knowledge if necessary. Two sessions with total length of 105 minutes are video recorded for post hoc analysis. We keep notes of the highlights with reference to their timeline position, flow of interaction, experts' interpretation, filter settings, as well as the video time. We report on our observation of cases in the rest of this section.
	
	\subsection
	{Checking Seasonal Distribution Change}
	\label{subsec:seasonal-distribution}
	\textbf{Background:} Seasonal climate change would impact many aspects in an ecosystem. Ecologists need to understand how this is reflected by animal movements. Fortunately, the raw data covers sufficient season cycles of multi-annual time frame. Thus, seasonal difference in movement distribution can be visually compared.
	
	\noindent\textbf{Method:} By either picking a specific position on the global timeline with mouse or manual input of time digits (\textbf{R2}), experts can quickly preview the general distribution of animal locations. Either way, the season (resp. southern hemisphere) display on the time ticker will change to the specified time. The experts also turn on the trace line to portray areas of denser movements by following two steps (\textbf{R1}): 1) they extend the trace line duration length to 90 days (meaning location history of roughly three months) and 2) they move the "current" time to the end of the season precisely by digits, e.g. "$28/02/2011~00:00$". Thus, a long trajectory would take on a nested shape within which condensed areas indicate frequent visits in particular regions. They can also filter out herbivores or predators to make clean comparisons between species across different time of the year (\textbf{R4}). Experts use the overall color tone of the ribbons in the chord diagram altogether to tell how close animals are forming (local) groups.
	
	\begin{figure}[h]
		\centering
		\includegraphics[width=\linewidth]{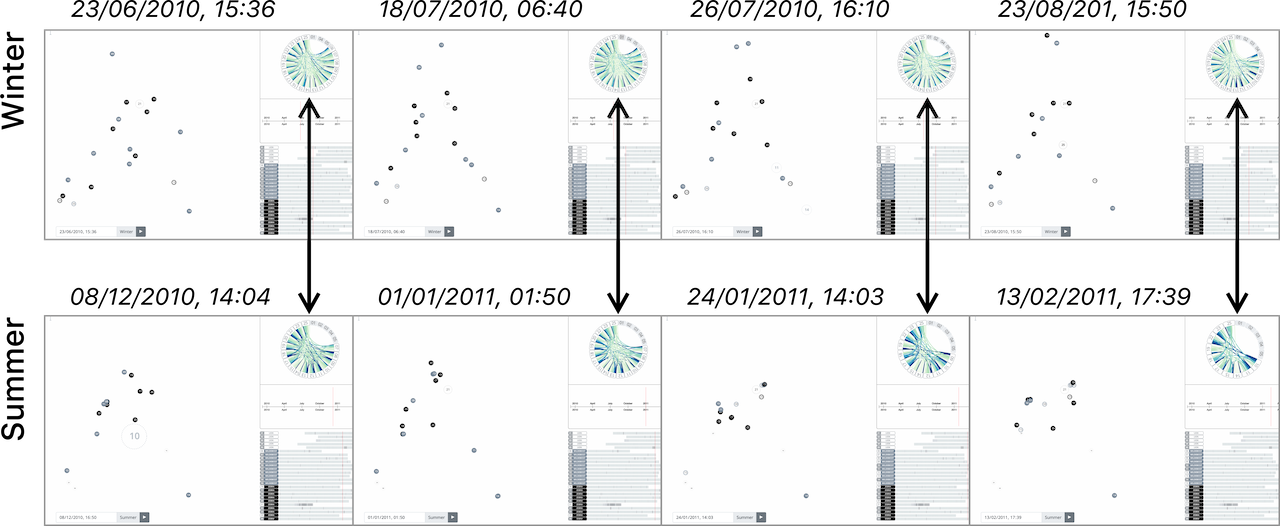}
		\caption{Herbivore spatial distribution and general relatedness level between summer and winter time. Chord diagram with more saturated ribbons indicates more related spatial distribution of entities.}
		\label{fig:seasonal-distribution-snapshots}
	\end{figure}
	\begin{figure}[h]
		\centering
		\includegraphics[width=.85\linewidth]{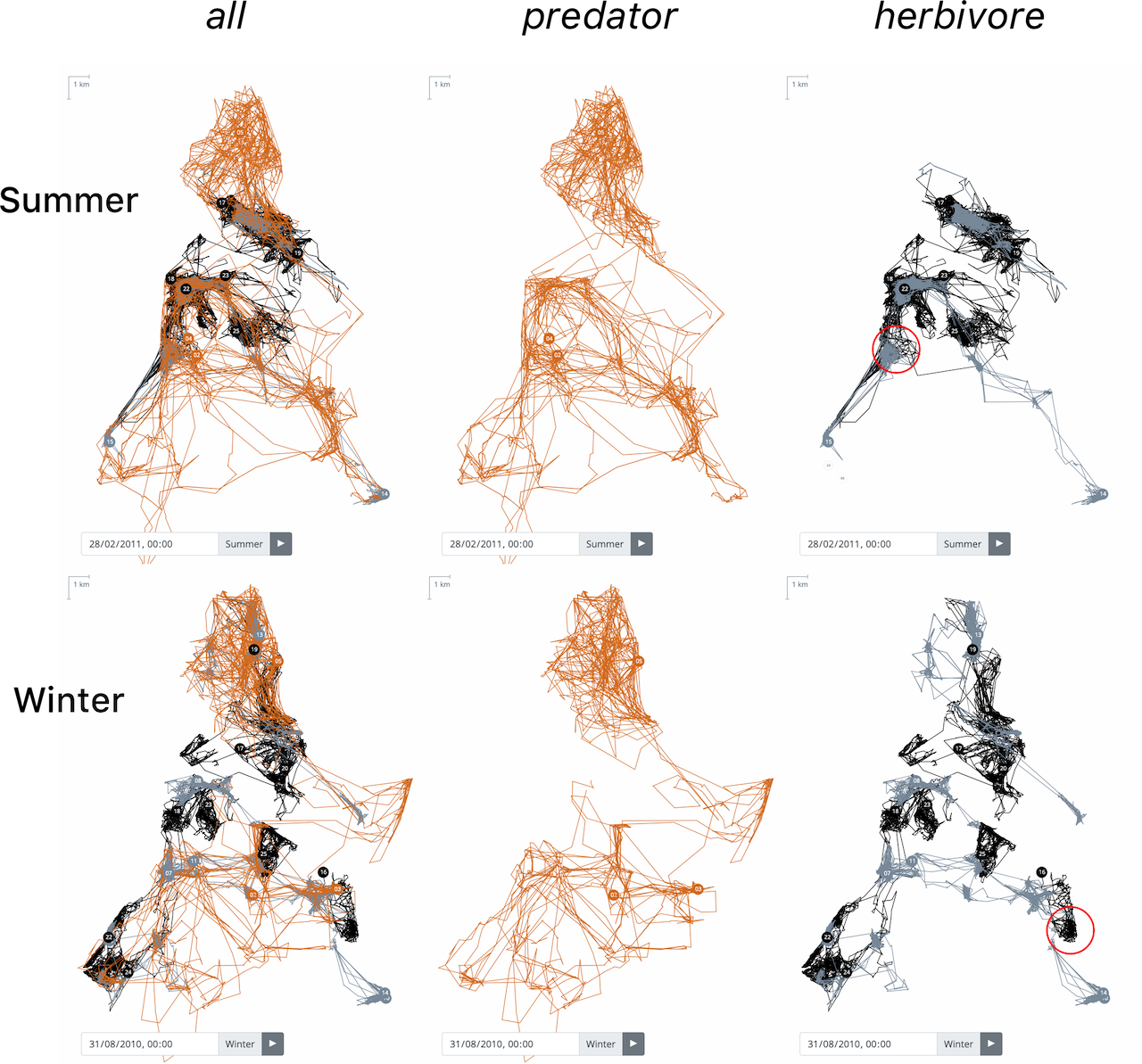}
		\caption{Long trajectory covering the movement paths of the entire season. Such configuration is useful to detect popular regions in a specific time of the year. Herbivores tend to concentrate in small patches in the center of the map during summer.}
		\label{fig:seasonal-movement-region}
	\end{figure}
	
	\noindent\textbf{Insight:}  \autoref{fig:seasonal-distribution-snapshots} are snapshots of animal distributions in winter vs. summer. It is observable that the herbivores tend to spread loosely in winter and gather closer during summer. Such a pattern tends to gravitate toward a few specific regions as it can be shown in the long term season trajectory \autoref{fig:seasonal-movement-region}. The experts believe that this could be caused by the periodic rainfall change in a year which leads to denser natural resource such as vegetation and water accumulations in some regions. The outcome shows that the concentration of natural resources attracts herbivores while such a trend is less obvious among the predators. Also, the shape and size (without smoothing) of trace lines can provide useful clues regarding the relative sedentary/active state of an animal: longer, more dispersed lines indicate more active behaviors in the area, while the more condensed, wiggling lines suggests more sedentary ones. From the comparisons between summer months and winter months, experts has discovered/confirmed some obvious hypothesis that correlate to seasonal changes ---  1) season does have effect on the concentration tendencies of animals, 2) such tendency is more significant among herbivores than predators and 3) relative sedentary states are most observed among herbivores during the winter.
	
	\begin{figure}[h]
		\centering
		\includegraphics[width=.9\linewidth]{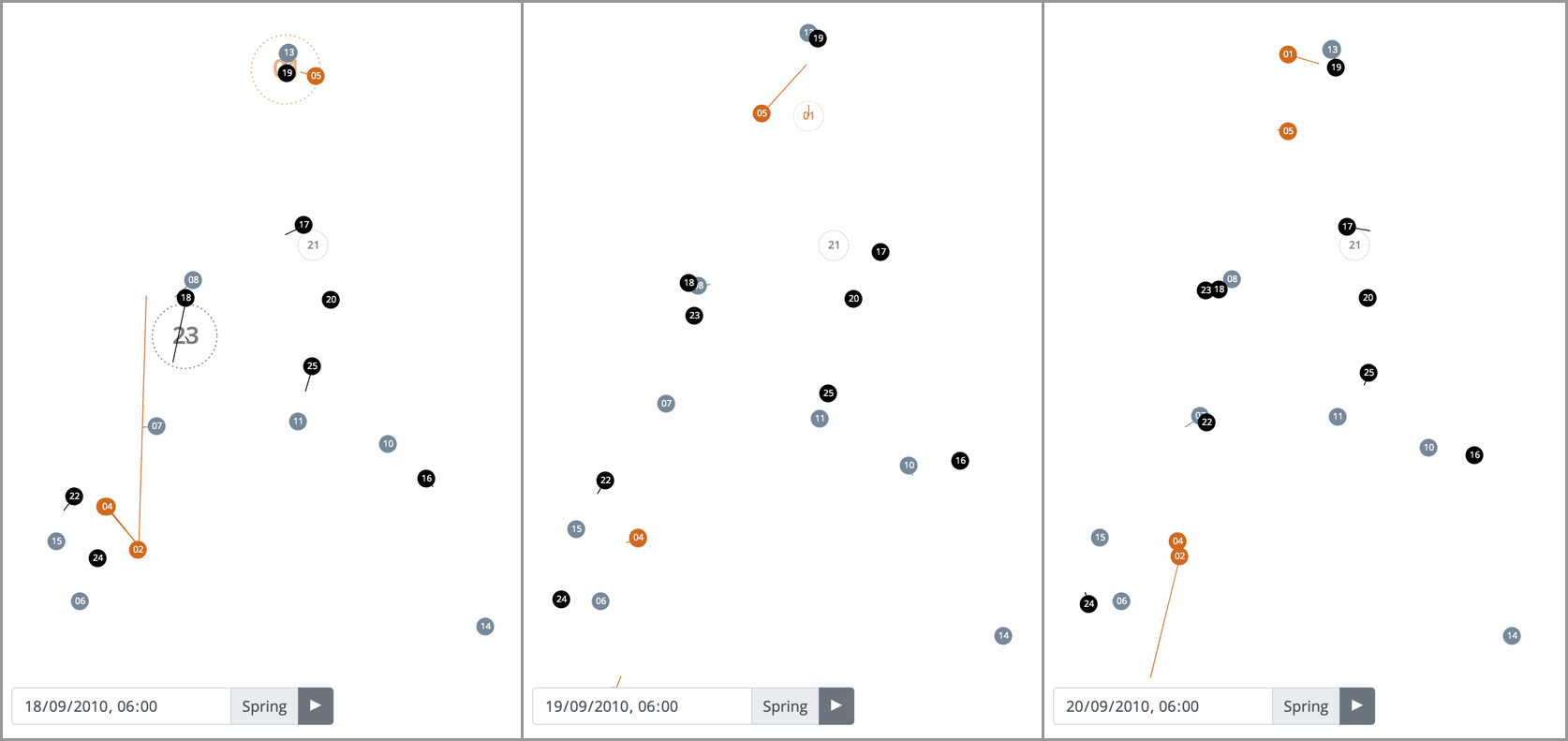}
		\caption{Absolute Travel Distance with Bundle Curve ($\alpha = 0$) : Predators travels much more during night hours.}
		\label{fig:night-activity}
	\end{figure}
	\subsection
	{Species Difference in Night Travel Distance}
	\textbf{Background:} Another potential influence on animal behavior is day-night transition. Unlike humans who tend to rest during most of the night hours, animals may react to day-night alternation differently in order to ensure their survival. The changes in temperature, visibility, as well as the effect of chronobiology could have heterogeneous effects on the movements of species.  How the behavior difference could be visually reflected by geographical patterns is intriguing.
	
	\noindent\textbf{Method:} The expert starts with setting the current time to 06:00 am on a random day and changes the trace line curve length to 9 hours. The main view displays movements trajectories of the last 9 hours --- from 09:00 pm the day before to 06:00 am current time. They can use arrow keys to jump to consecutive days without changing the time of day and trace line length. As a result, the common pattern in night movements are more directly exposed to the viewer (\textbf{R2}). As the shape of trace lines can be morphed with type and smoothness, the experts select "bundle" for curved type and slide the $\alpha$ value to maximum smoothing ($\alpha = 1$), which produces straight lines that connect only the origin and destination of the entire movement, cf. \autoref{fig:curve-parameters}. This aggressive smoothing technique allows the experts to make sense of the absolute travel distance during the night hours (\textbf{R1}).
	
	\noindent\textbf{Insight:} \autoref{fig:night-activity} shows an example of the observations by experts. They found little correlation between the night activities and geographical distributions. However, the night travel distances of predators (lions) are distinctively longer than ones from herbivores. This indicates that lions are more active during the nights while wildebeests and zebras tend to stay and rest as much as possible in the meanwhile.
	
	\subsection
	{Examining Grouping/Pairing Behavior}
	\textbf{Background:} Nuanced understanding of animal social interactions with an awareness of species traits plays a key role in the study of animal behaviors. As mentioned before, the grouping and pairing are difficult to detect with the visualization of mere locations due to spatial-temporal dependency. Instead, visualizing the dynamic relatedness can support the ecologists to investigate the strength of social bounding in pairs. 
	
	\noindent\textbf{Method:} When experts find that two animals are potentially forming a pair as they stay together drawing close, comparable traces, the experts select the corresponding animal pair in the dropdown menu from the control panel (\autoref{fig:overview}~C). The global pairwise relatedness is then plotted to delineate the intimacy between the two animals. Ups and downs that vary from month to month or season to season can be observed. As the expert brushes on the lower half of the chart around the current time, indicated by a vertical red line, fine variation of relatedness in the brushed zone are magnified to a daily or hourly level of detail for examination (\textbf{R5}). By checking the line shape of relatedness on both macro and micro level, experts can make a more reflective judgment on the grouping or pairing behavior.
	
	\begin{figure}[h]
		\centering
		\includegraphics[width=.9\linewidth]{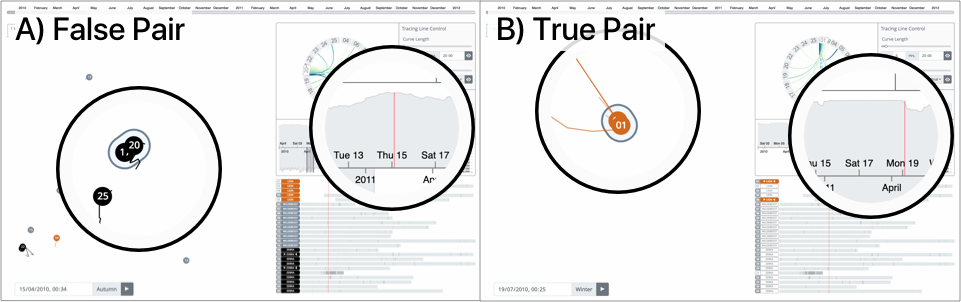}
		\caption{Pairing Examples: A) Zebras 20 and 17 seem to be socially close but their relatedness is not strong or stable enough along over time. B) Distance between lions 1 and 5 have diminished for roughly 2.5 km in the last 5 days 20 hours (filtered). Stable high pairwise relatedness is found repeatedly between the two lions. }
		\label{fig:true-false-pairing}
	\end{figure}
	
	\noindent\textbf{Insight:} The situation in A)~\autoref{fig:true-false-pairing} can be easily identified as pairing or grouping behavior by only looking at their temporary collocation from their trace line. The blue ribbon in the chord diagram that connects animal 17 and 20 seems to confirm that for the past 20 hours, the two are staying rather close together. But the pairwise relatedness view suggests that such relatedness is constantly changing and unstable. Thus, attributing grouping or pairing behavior is questionable in this case and needs further investigation. Another example in B)~\autoref{fig:true-false-pairing} tells a very different story of stable pairing ---  lions 1 and 5 have approached each other and maintained near maximum relatedness for a rather extended time. Since the phenomenon has repeated several times, experts believe the pattern is a more reliable indicator of a strong social pairing. Similar patterns have also been discovered between lions 2 and 4, wildebeests 12 and 13 using the same technique. A stable one-month pairing is also found between wildebeests 10 and 14. But the time window for this is too small comparing to the other groups and no further rejoining is found. More investigation might be needed to justify if the pairing is strong. The experts learn that a safer identification of intimate pairs requires evidence of close movements for longer period. Members of strong social pairings are likely to rejoin each other soon after separation. 
	
	\subsection{Analyzing  Multi-Species Relatedness}
	\label{subsec:analyzing-multi-species-relatedness}
	\begin{figure*}[h]
		\centering
		\includegraphics[width=1\linewidth]{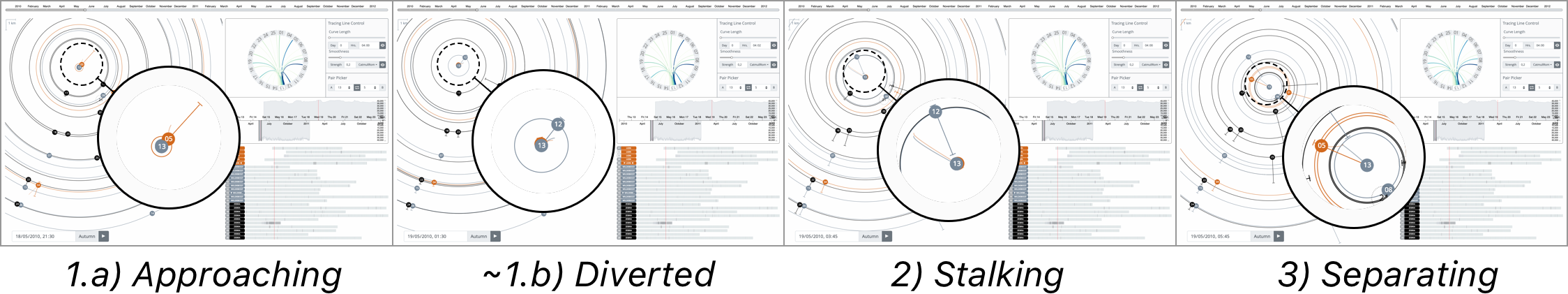}
		\caption{Three Stages of a Potential Encounter: 1.a) Increased relatedness between lion 5 and  group of wildebeest 9, 12, 13 suggests the movement pattern of one approaching a group of animals. 1.b) Wildebeest 12 has diverted from the group. 2) Stabled relatedness (diminishing capped line length for relatedness variance) shows the two species kept close distance during  the most of preceding 4 hours. 3) Lion 5 moved away accompanied by relatedness decrease. Both wildebeest 9 and 13 were found active after potential encounter.}
		\label{fig:predation-phases}
	\end{figure*}
	\textbf{Background:} Unlike relatedness within the same species, the inter-species relatedness may indicate threat instead of cooperation particularly between predators and herbivores. According to E1, a real predation process usually takes place within 3-5 minutes, which is beyond the frequency resolution of the employed GPS devices. However, unexplored behavioral patterns and multi-species interactions can still take place over the span of hours, according to E1. The experts would like to use inter-species relatedness to investigate possible instances of such behaviors.
	
	\noindent\textbf{Method:} Animals of more than one species can be seen as a potential stalking if one movement goes after another continuously. The smoothed trace line with duration length of several hours can be used to validate if the exact travel path fits well (\textbf{R1}). The expert uses the numerical input to fine-tune the trace line length to estimate the exact stalking time window. Clicking on one of the animals in the main view will trigger the i-G relatedness of between the selected animal and the distant one(s) of a different species B. This lets the experts knows the distance other animal(s) have covered to approach the selected one. In this case, the start and end of such episodes can be inferred with the help of pairwise relatedness.(\textbf{R5}).
	
	\noindent\textbf{Insight:} The example in \autoref{fig:predation-phases} illustrates how inter-species interactions can be studied through the spatial-temporal relatedness. The involved animals are lion 5 and wildebeests 9, 12, 13. Through each stage, the relatedness would witness an increase-stable-decrease process in the inter-species relatedness episode. The development of such a pattern is can be interpreted as hypothetical threat between predators and herbivores. But without concrete ground truth like direct observations, we can only assert the likelihood.
	
	\section
	{Evaluation}
	\label{sec:evaluation}
	We invited the experts (cf.~\secref{subsec:project-background}) to give conclusive remarks on the design. The evaluation starts with an open Q\&A session of 20 minutes each, during which we first clear up confusions on both sides, e.g. assumptions of animal behavior, or misinterpretation of visual variables. After that, each expert summarizes a final evaluation in written form. The result was collected by delivering questionnaires containing questions in three implicit facets to validate the design's usefulness -- the enabling, the facilitating, and the applicability. \textit{Enabling} emphasizes the aspects that provide novel analytical capabilities to find undiscovered insights whereas \textit{facilitating} consists of cases where the system solves their existing problems with a significant productivity boost. \textit{Applicability} addresses the conditions and contexts under which the system would achieve its maximum value. We set no strict time constrains to allow for as detailed answers as possible. The findings are given and discussed below.
	
	\subsection
	{Enabling} 
	According to E1, an animal can be influenced not only by variation in environmental conditions but also by the behavior and location of another individual or group of animals. Because of displacement in space and time, such relationships are tricky to explore visually. The visualization system allows them to analyze movement through space/time variation (\textbf{R1}) of individuals as well as the relationships between individuals (\textbf{R5}). The expert asserts that exploration into inter-individual interactions is enabled by the chord diagram with color-coded relatedness. 
	
	Regarding the pairwise relatedness, the ability to zoom to specific time periods (in pairwise relatedness) is very convenient and easy to use. E2 believes the visualization system highlighted an important capability which is visualizing data along temporal dimension, particularly how relatedness changes over time (\textbf{R2}).
	
	The i-G relatedness enables a visual understanding of the relatedness with actual distance between individuals in smaller time frames. The relatedness variation range raises the awareness of the time dependency in shorter movement trends. Unexpected patterns would emerge after testing and exploring with varying time scales (\textbf{R2}). E2 believes that the view mode is not only useful for generating ecological insights or hypotheses, but also creates more contextual awareness for the analysis. 
	
	\subsection
	{Facilitating}
	Before using our system, plotting static figures is their primary way of visual analysis for movements. According to their comments, the visualization creates depictions beyond static figures, without which the dynamics in movements are otherwise hard to interpret. "(Such functionality) is very needed in data exploration", says E1 (\textbf{R1}). 
	
	As they are fully aware of the difficulties introduced by inconsistent data, the new visual approach to check data availability/uncertainty is well-appreciated. "To me, this is a very useful tool for exploration of movement data, allowing to focus on different potential problems, such as sociality between individuals, movements relative to predators, home ranging, etc." says E1. The expert also confirms that the awareness of uncertainty is reinforced by trace lines and smoothing parameters which can be used to smooth out uncertain measurements (\textbf{R3}). E2 appreciates the quick configuration changes on the fly. He says, "It makes comparisons between configurations very convenient." (\textbf{R1}).
	
	\subsection{Applicability}
	Movement ecology research often requires calculating implicit features from the data such as road impact\cite{klar_effects_2009} or stigmergy\cite{giuggioli_stigmergy_2013}. When the optimal features to describe animal behaviors are still unclear and yet to be confirmed, the research becomes challenging. Based on their experience, the experts believe that the system can play a key role in their exploratory stage of analysis, where setting different parameters and scoping down to subsets of data need to be frequently adjusted. Under such circumstances, a comprehensive integration of capabilities that can produce easy to interpret visual insights with quick and convenient configuration changes is essential. According to E2, visualizing certain variables in a spatial-temporal way has changed their way of computing variables, doing analysis, and develop new hypotheses or insights.
	
	\section
	{Discussion}
	\label{sec:discussion}
	Comparing to other visualization works in supporting ecological research, this paper emphasizes more on individual level inquiries, particularly potential interactions or threats. The study into these issues initially came as a wicked problem that lacks a proper structure. 
	However, the perspective of relatedness puts the related information into clarified vision of movements in a relational context. This new attribute is not only a metric to aggregate proximity in time but also a variation pattern with a domain-friendly interpretation. In one way, exposing such patterns helps the expert to understand the unstable nature of spatial relation (cf. \autoref{fig:true-false-pairing}), which enables comparisons that lead to deeper, more intuitive understanding. In another way, the new attribute is a mathematical inference from the original data (cf.~\secref{subsec:defining-spatial-temporal-relatedness}), which keeps the possibility to revert the patterns to source values and (re)model or test if necessary.
	
	We acknowledge that distribution and behavior patterns caused by local resource change and land feature are as interesting to many ecologists, for example, the influence of seasonal change as in \secref{subsec:seasonal-distribution}. However, reasoning with more comprehensive factors such as temperature, rainfall is only feasible until we hold richer climate and GIS information of the area. Similarly, validation of visual interpretations such as predatory threat (cf.~\secref{subsec:analyzing-multi-species-relatedness}) relies on obtaining the ground truth data, for instance, recorded lion kills locations. Like before, the evidence is even more expensive to get based on current technology. In this regard, the visualization is helpful as it guides us toward how future research could be improved by obtaining additional knowledge and data.

	\section
	{Conclusion}
	\label{sec:conclusion}

	This article introduces a visual exploration system for movement ecologists with a focus on individual level insights into animal movements. 
	Although analyzing individual level interactions has been touched by previous works\cite{demsar_analysis_2015, andrienko_visual_2013-6}, a visualization tool that analyze small scale animal interactivity through the lens of relatedness is novel. Experts believe the design is useful in identifying of general movement patterns as well as locating possible pairing and matching in different time frames. The practicality of relatedness approach is supported by real cases and novel insights. Targeting at the emerging field of movement ecology\cite{nathan_emerging_2008, holden_inching_2006}, we see our work as a useful contribution to support nuanced insights in fine scale relationships of multi-species animals. 
	
	 In the future work, we are intrigued to find more supporting cases with species of different travel distances. For example, experts suggest the i-G relatedness approach to be viable in exploring predators’ prey base by searching for proximate herbivore candidates. Since our sample size is still small comparing to the number of all herbivores roaming in the area, redesigning the project setup by tracking a significant proportion of all possible preys of a different predator can be interesting. We also assume the usefulness of our methods in outlining shared behavioral principles by comparing human social movements with ones of wildlife. The insights could benefit the design of public spaces to promote social activities.
	
	
	
	\bibliographystyle{IEEEtran}
	\bibliography{refs}

	\begin{IEEEbiography}[{\includegraphics[width=1in,height=1.25in,clip,keepaspectratio]{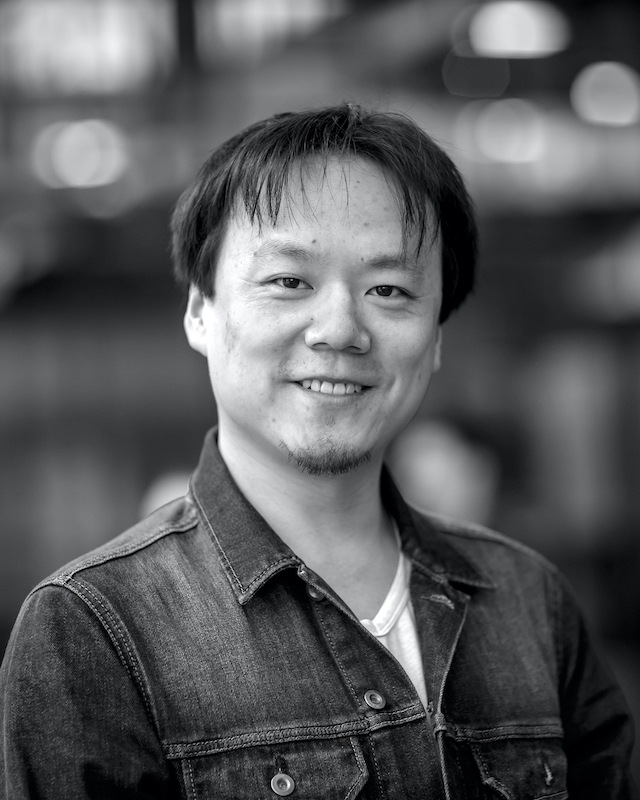}}]{Wei Li}
		received the MSc degree in Industrial Design Engineering from School of Design, Jiangnan University, China. He is currently a PhD candidate in the Department of Industrial Design, Eindhoven University of Technology, the Netherlands. His current research interests include visualization application, movement behavior, game analytics, and healthcare visualization.
	\end{IEEEbiography}
	\vspace*{-2\baselineskip}
	\begin{IEEEbiography}[{\includegraphics[width=1in,height=1.25in,clip,keepaspectratio]{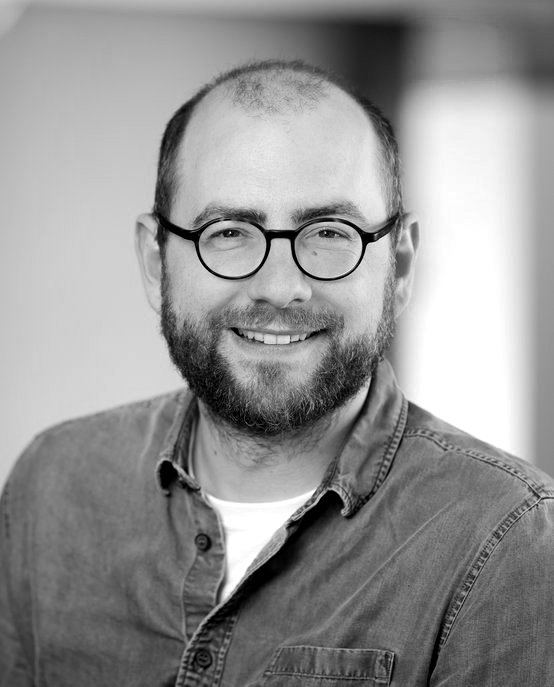}}]{Mathias Funk}
		has a background in Computer Science from RWTH Aachen, Germany, and got his PhD in Electrical Engineering at Eindhoven University of Technology. He is associate professor in the Department of Industrial Design, Eindhoven University of Technology. He leads the Things Ecology lab and is interested in design theory and processes for systems, designing systems for musical expression, and designing with data. 
	\end{IEEEbiography}
	\vspace*{-2\baselineskip}
	\begin{IEEEbiography}[{\includegraphics[width=1in,height=1.25in,clip,keepaspectratio]{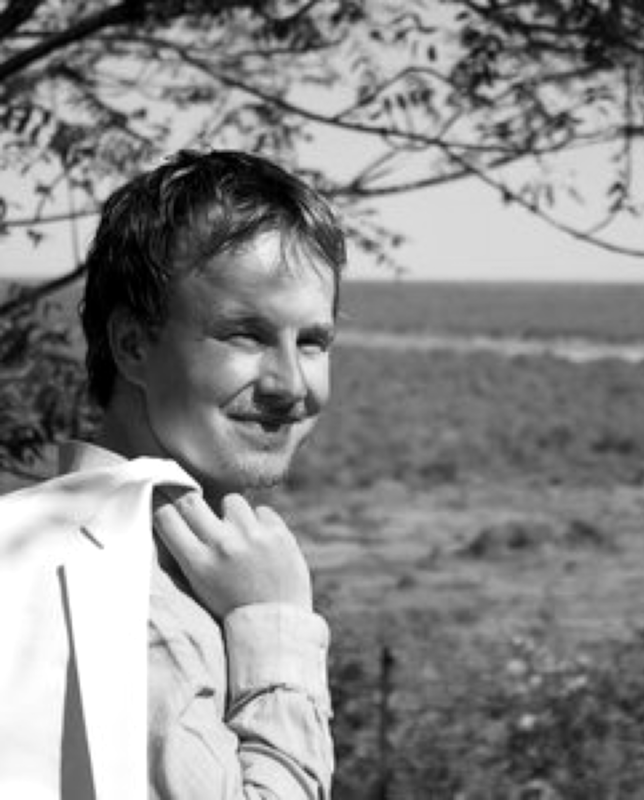}}]{Jasper  Eikelboom}
		  received received a MSc degree in Environmental Biology \& MA degree in Geography Education both at Utrecht University. He is currently PhD candidate in quantitative ecology at the Resource Ecology group of Wageningen University. His current research interests include movement ecology, machine learning, and conservation biology.
	\end{IEEEbiography}
	\vspace*{-2\baselineskip}
	\begin{IEEEbiography}[{\includegraphics[width=1in,height=1.25in,clip,keepaspectratio]{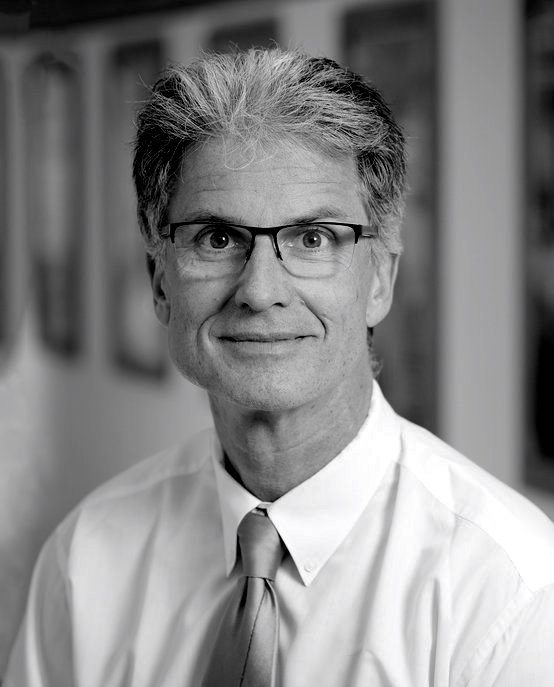}}]{Aarnout  Brombacher}
		holds a BSc and MSc in Electrical Engineering and a PhD in Engineering Science, all from Twente University of Technology. He was appointed full professor at Eindhoven University of Technology. He has also worked as a senior consultant for the Reliability section of Philips CFT Development Support. He is currently member of the National TopTeam (advisory body of the Dutch government) on Sports and Vitality representing the 14 Dutch universities in this field. He is interested in sports, activity for human health, and individual activity data.
	\end{IEEEbiography}
	\vfill
\end{document}

%% file: customization.tex
\newcommand{\etal}{\textit{~et~al.}}
\usepackage[T1]{fontenc}
\usepackage{amsmath}
\usepackage{hyperref}
\usepackage{cleveref}
\newcommand{\secref}[1]{\textit{\ref{#1} {\nameref{#1}}}}
\usepackage{amsfonts}
\usepackage{textcomp}
\usepackage{amssymb}
\usepackage{balance}

\newcommand{\out}[1]{}

%% file: _INDEX.bbl
\begin{thebibliography}{10}
\providecommand{\url}[1]{#1}
\csname url@samestyle\endcsname
\providecommand{\newblock}{\relax}
\providecommand{\bibinfo}[2]{#2}
\providecommand{\BIBentrySTDinterwordspacing}{\spaceskip=0pt\relax}
\providecommand{\BIBentryALTinterwordstretchfactor}{4}
\providecommand{\BIBentryALTinterwordspacing}{\spaceskip=\fontdimen2\font plus
\BIBentryALTinterwordstretchfactor\fontdimen3\font minus
  \fontdimen4\font\relax}
\providecommand{\BIBforeignlanguage}[2]{{%
\expandafter\ifx\csname l@#1\endcsname\relax
\typeout{** WARNING: IEEEtran.bst: No hyphenation pattern has been}%
\typeout{** loaded for the language `#1'. Using the pattern for}%
\typeout{** the default language instead.}%
\else
\language=\csname l@#1\endcsname
\fi
#2}}
\providecommand{\BIBdecl}{\relax}
\BIBdecl

\bibitem{holden_inching_2006}
C.~Holden, ``\BIBforeignlanguage{en}{Inching {{Toward Movement Ecology}}},''
  \emph{\BIBforeignlanguage{en}{Science}}, vol. 313, no. 5788, pp. 779--782,
  Aug. 2006.

\bibitem{cagnacci_managing_2008}
F.~Cagnacci and F.~Urbano, ``\BIBforeignlanguage{en}{Managing wildlife: {{A}}
  spatial information system for {{GPS}} collars data},''
  \emph{\BIBforeignlanguage{en}{Environmental Modelling \& Software}}, vol.~23,
  no.~7, pp. 957--959, Jul. 2008.

\bibitem{gor_gata_2017}
M.~Gor, J.~Vora, S.~Tanwar, S.~Tyagi, N.~Kumar, M.~S. Obaidat, and B.~Sadoun,
  ``{{GATA}}: {{GPS}}-{{Arduino}} based {{Tracking}} and {{Alarm}} system for
  protection of wildlife animals,'' in \emph{2017 {{International Conference}}
  on {{Computer}}, {{Information}} and {{Telecommunication Systems}}
  ({{CITS}})}.\hskip 1em plus 0.5em minus 0.4em\relax {Dalian, China}: {IEEE},
  Jul. 2017, pp. 166--170.

\bibitem{hoflinger_motion_2015}
F.~Hoflinger, R.~Zhang, T.~Volk, E.~{Garea-Rodriguez}, A.~Yousaf,
  C.~Schlumbohm, K.~Krieglstein, and L.~M. Reindl,
  ``\BIBforeignlanguage{en}{Motion capture sensor to monitor movement patterns
  in animal models of disease},'' in \emph{\BIBforeignlanguage{en}{2015
  {{IEEE}} 6th {{Latin American Symposium}} on {{Circuits}} \& {{Systems}}
  ({{LASCAS}})}}.\hskip 1em plus 0.5em minus 0.4em\relax {Montevideo, Uruguay}:
  {IEEE}, Feb. 2015, pp. 1--4.

\bibitem{qin_jiang_recognition_2004}
{Qin Jiang} and C.~Daniell, ``\BIBforeignlanguage{en}{Recognition of human and
  animal movement using infrared video streams},'' in
  \emph{\BIBforeignlanguage{en}{2004 {{International Conference}} on {{Image
  Processing}}, 2004. {{ICIP}} '04.}}, vol.~2.\hskip 1em plus 0.5em minus
  0.4em\relax {Singapore}: {IEEE}, 2004, pp. 1265--1268.

\bibitem{teimouri_deriving_2018}
M.~Teimouri, U.~Indahl, H.~Sickel, and H.~Tveite,
  ``\BIBforeignlanguage{en}{Deriving {{Animal Movement Behaviors Using Movement
  Parameters Extracted}} from {{Location Data}}},''
  \emph{\BIBforeignlanguage{en}{ISPRS International Journal of
  Geo-Information}}, vol.~7, no.~2, p.~78, Feb. 2018.

\bibitem{sarkar_analyzing_2015}
D.~Sarkar, C.~A. Chapman, L.~Griffin, and R.~Sengupta,
  ``\BIBforeignlanguage{en}{Analyzing {{Animal Movement Characteristics From
  Location Data}}: {{Analyzing Animal Movement}}},''
  \emph{\BIBforeignlanguage{en}{Transactions in GIS}}, vol.~19, no.~4, pp.
  516--534, Aug. 2015.

\bibitem{wang_new_2016}
Y.~Wang, Z.~Luo, J.~Takekawa, D.~Prosser, Y.~Xiong, S.~Newman, X.~Xiao,
  N.~Batbayar, K.~Spragens, S.~Balachandran, and B.~Yan,
  ``\BIBforeignlanguage{en}{A new method for discovering behavior patterns
  among animal movements},'' \emph{\BIBforeignlanguage{en}{International
  Journal of Geographical Information Science}}, vol.~30, no.~5, pp. 929--947,
  May 2016.

\bibitem{li_mining_2012}
Z.~Li, J.~Han, B.~Ding, and R.~Kays, ``\BIBforeignlanguage{en}{Mining periodic
  behaviors of object movements for animal and biological sustainability
  studies},'' \emph{\BIBforeignlanguage{en}{Data Mining and Knowledge
  Discovery}}, vol.~24, no.~2, pp. 355--386, Mar. 2012.

\bibitem{li_movemine_2011}
Z.~Li, J.~Han, M.~Ji, L.-A. Tang, Y.~Yu, B.~Ding, J.-G. Lee, and R.~Kays,
  ``\BIBforeignlanguage{en}{{{MoveMine}}: {{Mining}} moving object data for
  discovery of animal movement patterns},'' \emph{\BIBforeignlanguage{en}{ACM
  Transactions on Intelligent Systems and Technology}}, vol.~2, no.~4, pp.
  1--32, Jul. 2011.

\bibitem{nathan_movement_2008}
R.~Nathan, W.~M. Getz, E.~Revilla, M.~Holyoak, R.~Kadmon, D.~Saltz, and P.~E.
  Smouse, ``A movement ecology paradigm for unifying organismal movement
  research,'' \emph{Proceedings of the National Academy of Sciences}, vol. 105,
  no.~49, pp. 19\,052--19\,059, 2008.

\bibitem{kranstauber_movebank_2011}
B.~Kranstauber, A.~Cameron, R.~Weinzerl, T.~Fountain, S.~Tilak, M.~Wikelski,
  and R.~Kays, ``The {{Movebank}} data model for animal tracking,''
  \emph{Environmental Modelling \& Software}, vol.~26, no.~6, pp. 834--835,
  Jun. 2011.

\bibitem{cagnacci_animal_2010}
F.~Cagnacci, L.~Boitani, R.~Powell, and M.~Boyce, ``Animal ecology meets
  {{GPS}}-based radiotelemetry: {{A}} perfect storm of opportunities and
  challenges,'' \emph{Philosophical Transactions of the Royal Society B:
  Biological Sciences}, vol. 365, no. 1550, pp. 2157--2162, 2010.

\bibitem{spretke_exploration_2011}
D.~Spretke, P.~Bak, H.~Janetzko, B.~Kranstauber, F.~Mansmann, and S.~Davidson,
  ``Exploration {{Through Enrichment}}: {{A Visual Analytics Approach}} for
  {{Animal Movement}},'' in \emph{Proceedings of the 19th {{ACM SIGSPATIAL
  International Conference}} on {{Advances}} in {{Geographic Information
  Systems}}}, ser. {{GIS}} '11.\hskip 1em plus 0.5em minus 0.4em\relax {New
  York, NY, USA}: {ACM}, 2011, pp. 421--424.

\bibitem{nathan_emerging_2008}
R.~Nathan, ``\BIBforeignlanguage{en}{An emerging movement ecology paradigm},''
  \emph{\BIBforeignlanguage{en}{Proceedings of the National Academy of
  Sciences}}, vol. 105, no.~49, pp. 19\,050--19\,051, Dec. 2008.

\bibitem{westley_peter_a._h._collective_2018}
{Westley Peter A. H.}, {Berdahl Andrew M.}, {Torney Colin J.}, and {Biro Dora},
  ``Collective movement in ecology: From emerging technologies to conservation
  and management,'' \emph{Philosophical Transactions of the Royal Society B:
  Biological Sciences}, vol. 373, no. 1746, p. 20170004, May 2018.

\bibitem{holyoak_trends_2008}
M.~Holyoak, R.~Casagrandi, R.~Nathan, E.~Revilla, and O.~Spiegel,
  ``\BIBforeignlanguage{en}{Trends and missing parts in the study of movement
  ecology},'' \emph{\BIBforeignlanguage{en}{Proceedings of the National Academy
  of Sciences}}, vol. 105, no.~49, pp. 19\,060--19\,065, Dec. 2008.

\bibitem{calabrese_disentangling_2018}
J.~M. Calabrese, C.~H. Fleming, W.~F. Fagan, M.~Rimmler, P.~Kaczensky,
  S.~Bewick, P.~Leimgruber, and T.~Mueller,
  ``\BIBforeignlanguage{en}{Disentangling social interactions and environmental
  drivers in multi-individual wildlife tracking data},''
  \emph{\BIBforeignlanguage{en}{Philosophical Transactions of the Royal Society
  B: Biological Sciences}}, vol. 373, no. 1746, p. 20170007, May 2018.

\bibitem{giuggioli_stigmergy_2013}
L.~Giuggioli, J.~R. Potts, D.~I. Rubenstein, and S.~A. Levin,
  ``\BIBforeignlanguage{en}{Stigmergy, collective actions, and animal social
  spacing},'' \emph{\BIBforeignlanguage{en}{Proceedings of the National Academy
  of Sciences}}, vol. 110, no.~42, pp. 16\,904--16\,909, Oct. 2013.

\bibitem{polansky_framework_2011}
L.~Polansky and G.~Wittemyer, ``\BIBforeignlanguage{en}{A framework for
  understanding the architecture of collective movements using pairwise
  analyses of animal movement data},'' \emph{\BIBforeignlanguage{en}{Journal of
  The Royal Society Interface}}, vol.~8, no.~56, pp. 322--333, Mar. 2011.

\bibitem{strandburg-peshkin_inferring_2018}
A.~{Strandburg-Peshkin}, D.~Papageorgiou, M.~C. Crofoot, and D.~R. Farine,
  ``\BIBforeignlanguage{en}{Inferring influence and leadership in moving animal
  groups},'' \emph{\BIBforeignlanguage{en}{Philosophical Transactions of the
  Royal Society B: Biological Sciences}}, vol. 373, no. 1746, p. 20170006, May
  2018.

\bibitem{torney_inferring_2018}
C.~J. Torney, M.~Lamont, L.~Debell, R.~J. Angohiatok, L.-M. Leclerc, and A.~M.
  Berdahl, ``\BIBforeignlanguage{en}{Inferring the rules of social interaction
  in migrating caribou},'' \emph{\BIBforeignlanguage{en}{Philosophical
  Transactions of the Royal Society B: Biological Sciences}}, vol. 373, no.
  1746, p. 20170385, May 2018.

\bibitem{pires_interaction_2012}
M.~M. Pires and P.~R. Guimaraes, ``\BIBforeignlanguage{en}{Interaction intimacy
  organizes networks of antagonistic interactions in different ways},''
  \emph{\BIBforeignlanguage{en}{Journal of The Royal Society Interface}},
  vol.~10, no.~78, pp. 20\,120\,649--20\,120\,649, Nov. 2012.

\bibitem{hagen_biodiversity_2012}
M.~Hagen, W.~D. Kissling, C.~Rasmussen, M.~A. De~Aguiar, L.~E. Brown, D.~W.
  Carstensen, I.~{Alves-Dos-Santos}, Y.~L. Dupont, F.~K. Edwards, J.~Genini,
  P.~R. Guimar{\~a}es, G.~B. Jenkins, P.~Jordano, C.~N. {Kaiser-Bunbury}, M.~E.
  Ledger, K.~P. Maia, F.~M.~D. Marquitti, {\'O}.~Mclaughlin, L.~P.~C.
  Morellato, E.~J. O'Gorman, K.~Tr{\o}jelsgaard, J.~M. Tylianakis, M.~M. Vidal,
  G.~Woodward, and J.~M. Olesen, ``\BIBforeignlanguage{en}{Biodiversity,
  {{Species Interactions}} and {{Ecological Networks}} in a {{Fragmented
  World}}},'' in \emph{\BIBforeignlanguage{en}{Advances in {{Ecological
  Research}}}}.\hskip 1em plus 0.5em minus 0.4em\relax {Elsevier}, 2012,
  vol.~46, pp. 89--210.

\bibitem{slingsby_exploratory_2016}
A.~Slingsby and E.~{van Loon}, ``\BIBforeignlanguage{en}{Exploratory {{Visual
  Analysis}} for {{Animal Movement Ecology}}},''
  \emph{\BIBforeignlanguage{en}{Computer Graphics Forum}}, vol.~35, no.~3, pp.
  471--480, Jun. 2016.

\bibitem{eagle_inferring_2009}
N.~Eagle, A.~S. Pentland, and D.~Lazer, ``\BIBforeignlanguage{en}{Inferring
  friendship network structure by using mobile phone data},''
  \emph{\BIBforeignlanguage{en}{Proceedings of the National Academy of
  Sciences}}, vol. 106, no.~36, pp. 15\,274--15\,278, Sep. 2009.

\bibitem{roshier_animal_2008}
D.~A. Roshier, V.~A.~J. Doerr, and E.~D. Doerr,
  ``\BIBforeignlanguage{en}{Animal movement in dynamic landscapes: Interaction
  between behavioural strategies and resource distributions},''
  \emph{\BIBforeignlanguage{en}{Oecologia}}, vol. 156, no.~2, pp. 465--477, May
  2008.

\bibitem{ferreira_birdvis_2011}
N.~Ferreira, L.~Lins, D.~Fink, S.~Kelling, C.~Wood, J.~Freire, and C.~Silva,
  ``{{BirdVis}}: {{Visualizing}} and {{Understanding Bird Populations}},''
  \emph{IEEE Transactions on Visualization and Computer Graphics}, vol.~17,
  no.~12, pp. 2374--2383, Dec. 2011.

\bibitem{kavathekar_introducing_2013}
D.~Kavathekar, T.~Mueller, and W.~F. Fagan, ``Introducing {{AMV}} ({{Animal
  Movement Visualizer}}), a visualization tool for animal movement data from
  satellite collars and radiotelemetry,'' \emph{Ecological Informatics},
  vol.~15, pp. 91--95, May 2013.

\bibitem{dodge_environmental-data_2013}
S.~Dodge, G.~Bohrer, R.~Weinzierl, S.~C. Davidson, R.~Kays, D.~Douglas,
  S.~Cruz, J.~Han, D.~Brandes, and M.~Wikelski, ``The environmental-data
  automated track annotation ({{Env}}-{{DATA}}) system: Linking animal tracks
  with environmental data,'' \emph{Movement Ecology}, vol.~1, p.~3, Jul. 2013.

\bibitem{seebacher_visual_2018}
D.~Seebacher, J.~Haualer, M.~Hundt, M.~Stein, H.~Muller, U.~Engelke, and
  D.~Keim, ``\BIBforeignlanguage{en}{Visual {{Analysis}} of
  {{Spatio}}-{{Temporal Event Predictions}}: {{Investigating}} the {{Spread
  Dynamics}} of {{Invasive Species}}},'' \emph{\BIBforeignlanguage{en}{IEEE
  Transactions on Big Data}}, pp. 1--1, 2018.

\bibitem{xavier_exploratory_2014}
G.~Xavier and S.~Dodge, ``An {{Exploratory Visualization Tool}} for {{Mapping}}
  the {{Relationships Between Animal Movement}} and the {{Environment}},'' in
  \emph{Proceedings of the 22nd {{ACM SIGSPATIAL International Workshop}} on
  {{Interacting}} with {{Maps}}}, ser. {{MapInteract}} '14.\hskip 1em plus
  0.5em minus 0.4em\relax {New York, NY, USA}: {ACM}, 2014, pp. 36--42.

\bibitem{taylor_connectivity_1993}
P.~D. Taylor, L.~Fahrig, K.~Henein, and G.~Merriam, ``Connectivity {{Is}} a
  {{Vital Element}} of {{Landscape Structure}},'' \emph{Oikos}, vol.~68, no.~3,
  p. 571, Dec. 1993.

\bibitem{baguette_landscape_2007}
M.~Baguette and H.~Van~Dyck, ``\BIBforeignlanguage{en}{Landscape connectivity
  and animal behavior: Functional grain as a key determinant for dispersal},''
  \emph{\BIBforeignlanguage{en}{Landscape Ecology}}, vol.~22, no.~8, pp.
  1117--1129, Oct. 2007.

\bibitem{lima_towards_1996}
S.~L. Lima and P.~A. Zollner, ``\BIBforeignlanguage{en}{Towards a behavioral
  ecology of ecological landscapes},'' \emph{\BIBforeignlanguage{en}{Trends in
  Ecology \& Evolution}}, vol.~11, no.~3, pp. 131--135, Mar. 1996.

\bibitem{konzack_visual_2018}
M.~Konzack, P.~Gijsbers, F.~Timmers, E.~{van Loon}, M.~A. Westenberg, and
  K.~Buchin, ``\BIBforeignlanguage{en}{Visual exploration of migration patterns
  in gull data},'' \emph{\BIBforeignlanguage{en}{Information Visualization}},
  p. 147387161775124, Jan. 2018.

\bibitem{bak_scalable_2012}
P.~Bak, M.~Marder, S.~Harary, A.~Yaeli, and H.~J. Ship,
  ``\BIBforeignlanguage{en}{Scalable {{Detection}} of {{Spatiotemporal
  Encounters}} in {{Historical Movement Data}}},''
  \emph{\BIBforeignlanguage{en}{Computer Graphics Forum}}, vol.~31, no. 3pt1,
  pp. 915--924, 2012.

\bibitem{andrienko_event-based_2011}
G.~Andrienko, N.~Andrienko, and M.~Heurich, ``An event-based conceptual model
  for context-aware movement analysis,'' \emph{International Journal of
  Geographical Information Science}, vol.~25, no.~9, pp. 1347--1370, Sep. 2011.

\bibitem{siqueira_discovering_2011}
F.~d.~L. Siqueira and V.~Bogorny, ``\BIBforeignlanguage{en}{Discovering
  {{Chasing Behavior}} in {{Moving Object Trajectories}}},''
  \emph{\BIBforeignlanguage{en}{Transactions in GIS}}, vol.~15, no.~5, pp.
  667--688, 2011.

\bibitem{andrienko_uncovering_2008}
N.~V. Andrienko, G.~Andrienko, M.~Wachowicz, and D.~Orellana, ``Uncovering
  {{Interactions}} between {{Moving Objects}},'' 2008.

\bibitem{bertin_semiology_2010}
J.~Bertin, \emph{\BIBforeignlanguage{English}{Semiology of {{Graphics}}:
  {{Diagrams}}, {{Networks}}, {{Maps}}}}, 1st~ed.\hskip 1em plus 0.5em minus
  0.4em\relax {Redlands, Calif}: {Esri Press}, Nov. 2010.

\bibitem{shamoun-baranes_analysis_2012}
J.~{Shamoun-Baranes}, E.~E. van Loon, R.~S. Purves, B.~Speckmann, D.~Weiskopf,
  and C.~J. Camphuysen, ``\BIBforeignlanguage{en}{Analysis and visualization of
  animal movement},'' \emph{\BIBforeignlanguage{en}{Biology Letters}}, vol.~8,
  no.~1, pp. 6--9, Feb. 2012.

\bibitem{hedley_hagerstrand_1999}
N.~R. Hedley, C.~H. Drew, E.~A. Arfin, and A.~Lee, ``Hagerstrand {{Revisited}}:
  {{Interactive Space}}-{{Time Visualizations}} of {{Complex Spatial Data}},''
  \emph{Informatica (Slovenia)}, vol.~23, no.~2, 1999.

\bibitem{gatalsky_interactive_2004}
P.~Gatalsky, N.~Andrienko, and G.~Andrienko,
  ``\BIBforeignlanguage{en}{Interactive analysis of event data using space-time
  cube},'' in \emph{\BIBforeignlanguage{en}{Proceedings. {{Eighth International
  Conference}} on {{Information Visualisation}}, 2004. {{IV}} 2004.}}\hskip 1em
  plus 0.5em minus 0.4em\relax {London, England}: {IEEE}, 2004, pp. 145--152.

\bibitem{kraak_space_2003}
M.~Kraak, ``\BIBforeignlanguage{English}{The space - time cube revisited from a
  geovisualization perspective},'' in
  \emph{\BIBforeignlanguage{English}{{{ICC}} 2003 : {{Proceedings}} of the 21st
  International Cartographic Conference}}.\hskip 1em plus 0.5em minus
  0.4em\relax {International Cartographic Association (ICA)}, 2003, pp.
  1988--1996.

\bibitem{walsh_temporal-geospatial_2016}
J.~A. Walsh, J.~Zucco, R.~T. Smith, and B.~H. Thomas,
  ``\BIBforeignlanguage{en}{Temporal-{{Geospatial Cooperative Visual
  Analysis}}},'' in \emph{\BIBforeignlanguage{en}{2016 {{Big Data Visual
  Analytics}} ({{BDVA}})}}.\hskip 1em plus 0.5em minus 0.4em\relax {Sydney,
  Australia}: {IEEE}, Nov. 2016, pp. 1--8.

\bibitem{amini_impact_2015}
F.~Amini, S.~Rufiange, Z.~Hossain, Q.~Ventura, P.~Irani, and M.~J. McGuffin,
  ``\BIBforeignlanguage{en}{The {{Impact}} of {{Interactivity}} on
  {{Comprehending 2D}} and {{3D Visualizations}} of {{Movement Data}}},''
  \emph{\BIBforeignlanguage{en}{IEEE Transactions on Visualization and Computer
  Graphics}}, vol.~21, no.~1, pp. 122--135, Jan. 2015.

\bibitem{andrienko_clustering_2018}
G.~Andrienko, N.~Andrienko, G.~Fuchs, and J.~M.~C. Garcia, ``Clustering
  {{Trajectories}} by {{Relevant Parts}} for {{Air Traffic Analysis}},''
  \emph{IEEE Transactions on Visualization and Computer Graphics}, vol.~24,
  no.~1, pp. 34--44, Jan. 2018.

\bibitem{crnovrsanin_proximity-based_2009}
T.~Crnovrsanin, C.~Muelder, C.~Correa, and K.~L. Ma, ``Proximity-based
  visualization of movement trace data,'' in \emph{2009 {{IEEE Symposium}} on
  {{Visual Analytics Science}} and {{Technology}}}, Oct. 2009, pp. 11--18.

\bibitem{andrienko_visual_2013-6}
G.~Andrienko, N.~Andrienko, P.~Bak, D.~Keim, and S.~Wrobel,
  \emph{\BIBforeignlanguage{en}{Visual {{Analytics}} of {{Movement}}}}.\hskip
  1em plus 0.5em minus 0.4em\relax {Berlin, Heidelberg}: {Springer Berlin
  Heidelberg}, 2013.

\bibitem{handcock_monitoring_2009}
R.~Handcock, D.~Swain, G.~{Bishop-Hurley}, K.~Patison, T.~Wark, P.~Valencia,
  P.~Corke, and C.~O'Neill, ``\BIBforeignlanguage{en}{Monitoring {{Animal
  Behaviour}} and {{Environmental Interactions Using Wireless Sensor
  Networks}}, {{GPS Collars}} and {{Satellite Remote Sensing}}},''
  \emph{\BIBforeignlanguage{en}{Sensors}}, vol.~9, no.~5, pp. 3586--3603, May
  2009.

\bibitem{hurford_gps_2009}
A.~Hurford, ``\BIBforeignlanguage{en}{{{GPS Measurement Error Gives Rise}} to
  {{Spurious}} 180\textdegree{} {{Turning Angles}} and {{Strong Directional
  Biases}} in {{Animal Movement Data}}},'' \emph{\BIBforeignlanguage{en}{PLoS
  ONE}}, vol.~4, no.~5, p. e5632, May 2009.

\bibitem{bjorneraas_screening_2010}
K.~Bj{\o}rneraas, B.~Moorter, C.~M. Rolandsen, and I.~Herfindal,
  ``\BIBforeignlanguage{en}{Screening {{Global Positioning System Location
  Data}} for {{Errors Using Animal Movement Characteristics}}},''
  \emph{\BIBforeignlanguage{en}{The Journal of Wildlife Management}}, vol.~74,
  no.~6, pp. 1361--1366, Aug. 2010.

\bibitem{reas_processing_2007}
C.~Reas and B.~Fry, \emph{\BIBforeignlanguage{en}{Processing: A Programming
  Handbook for Visual Designers and Artists}}.\hskip 1em plus 0.5em minus
  0.4em\relax {Cambridge, Mass.}: {MIT Press}, 2007.

\bibitem{sacha_dynamic_2017}
D.~Sacha, F.~{Al-Masoudi}, M.~Stein, T.~Schreck, D.~A. Keim, G.~Andrienko, and
  H.~Janetzko, ``\BIBforeignlanguage{en}{Dynamic {{Visual Abstraction}} of
  {{Soccer Movement}}},'' \emph{\BIBforeignlanguage{en}{Computer Graphics
  Forum}}, vol.~36, no.~3, pp. 305--315, Jun. 2017.

\bibitem{demsar_analysis_2015}
U.~Dem{\v s}ar, K.~Buchin, F.~Cagnacci, K.~Safi, B.~Speckmann, N.~{Van de
  Weghe}, D.~Weiskopf, and R.~Weibel, ``Analysis and visualisation of movement:
  An interdisciplinary review,'' \emph{Movement Ecology}, vol.~3, p.~5, Mar.
  2015.

\bibitem{andrienko_space_2010}
G.~Andrienko, N.~Andrienko, U.~Demsar, D.~Dransch, J.~Dykes, S.~I. Fabrikant,
  M.~Jern, M.-J. Kraak, H.~Schumann, and C.~Tominski, ``Space, time and visual
  analytics,'' \emph{International Journal of Geographical Information
  Science}, vol.~24, no.~10, pp. 1577--1600, Oct. 2010.

\bibitem{riveiro_effects_2014}
M.~Riveiro, T.~Helldin, G.~Falkman, and M.~Lebram, ``Effects of visualizing
  uncertainty on decision-making in a target identification scenario,''
  \emph{Computers \& Graphics}, vol.~41, pp. 84--98, Jun. 2014.

\bibitem{wunderlich_visualization_2017}
M.~Wunderlich, K.~Ballweg, G.~Fuchs, and T.~{von Landesberger},
  ``\BIBforeignlanguage{en}{Visualization of {{Delay Uncertainty}} and its
  {{Impact}} on {{Train Trip Planning}}: {{A Design Study}}},''
  \emph{\BIBforeignlanguage{en}{Computer Graphics Forum}}, vol.~36, no.~3, pp.
  317--328, Jun. 2017.

\bibitem{sacha_role_2016}
D.~Sacha, H.~Senaratne, B.~C. Kwon, G.~Ellis, and D.~A. Keim, ``The {{Role}} of
  {{Uncertainty}}, {{Awareness}}, and {{Trust}} in {{Visual Analytics}},''
  \emph{IEEE Transactions on Visualization and Computer Graphics}, vol.~22,
  no.~1, pp. 240--249, Jan. 2016.

\bibitem{holten_hierarchical_2006}
D.~Holten, ``Hierarchical {{Edge Bundles}}: {{Visualization}} of {{Adjacency
  Relations}} in {{Hierarchical Data}},'' \emph{IEEE Transactions on
  Visualization and Computer Graphics}, vol.~12, no.~5, pp. 741--748, Sep.
  2006.

\bibitem{cockburn_review_2008}
A.~Cockburn, A.~Karlson, and B.~B. Bederson, ``\BIBforeignlanguage{en}{A review
  of overview+detail, zooming, and focus+context interfaces},''
  \emph{\BIBforeignlanguage{en}{ACM Computing Surveys}}, vol.~41, no.~1, pp.
  1--31, Dec. 2008.

\bibitem{boren_thinking_2000}
T.~Boren and J.~Ramey, ``Thinking aloud: Reconciling theory and practice,''
  \emph{IEEE Transactions on Professional Communication}, vol.~43, no.~3, pp.
  261--278, Sep. 2000.

\bibitem{klar_effects_2009}
N.~Klar, M.~Herrmann, and S.~{Kramer-Schadt}, ``\BIBforeignlanguage{en}{Effects
  and {{Mitigation}} of {{Road Impacts}} on {{Individual Movement Behavior}} of
  {{Wildcats}}},'' \emph{\BIBforeignlanguage{en}{Journal of Wildlife
  Management}}, vol.~73, no.~5, pp. 631--638, Jul. 2009.

\end{thebibliography}
